\documentclass[11pt, aps, preprint, nofootinbib,superscriptaddress,eqsecnum,titlepage]{revtex4-2}

\usepackage{array}[=2016-10-06]

\usepackage[active]{srcltx}
\usepackage[utf8]{inputenc}
\usepackage{enumerate}
\usepackage[normalem]{ulem}
\usepackage{manfnt}
\usepackage{physics}
\usepackage{latexsym}
\usepackage{hyperref}
\usepackage{amsmath}
\usepackage{amsfonts}
\usepackage{graphicx}
\usepackage{slashed}
\usepackage{xcolor}
\usepackage{soul}
\usepackage{slashed}
\usepackage{braket}
\usepackage{natbib}
\usepackage{cancel}
\usepackage{MnSymbol}
\usepackage{wasysym}
\usepackage[export]{adjustbox}
\usepackage{graphicx}
\usepackage{mathtools}

\usepackage{tabularx}
\newcolumntype{C}[1]{>{\centering\arraybackslash}p{#1}}
\newcolumntype{Y}{>{\centering\arraybackslash}X}

\usepackage{subcaption}

\usepackage[dvipsnames]{xcolor}

\definecolor{azure(colorwheel)}{rgb}{0.0, 0.5, 1.0}

\definecolor{crimsonglory}{rgb}{0.75, 0.0, 0.2}

\allowdisplaybreaks

\begin{document}
	
\title{\Large{{\sc Gauging Modulated Symmetries: Bond Algebras, Higher-Form Symmetries, and Symmetry-Enriched Topological Order}}}

\author{Gustavo M. Yoshitome}
\email{gustavoyoshitome@gmail.com}
\affiliation{Departamento de Física, Universidade Estadual de Londrina, Londrina, PR, Brasil}

\author{Pedro R. S. Gomes}
\email{pedrogomes@uel.br}
\affiliation{Departamento de Física, Universidade Estadual de Londrina, Londrina, PR, Brasil}

\author{Guilherme Delfino}
\email{gdelfino@purdue.edu}
\affiliation{Department of Physics and Astronomy, Purdue University, West Lafayette, Indiana, 47907, USA}

	
\begin{abstract}

We develop a constructive framework for characterizing modulated symmetries and their gauging in $d$-dimensional lattice systems. The modulation profiles are organized as a finite-dimensional representation of the space group acting on a finite Abelian internal symmetry group. Since the corresponding automorphism group is finite, the nontrivial spatial action factors through a finite quotient of the space group, leading to periodic modulation patterns and providing a classification framework for discrete Abelian modulated symmetries. The symmetry-allowed local interactions are systematically obtained from a kernel construction, which also yields an algebra of open operators. Upon gauging, its elements are transmuted into extended Wilson and boundary operators while retaining their algebraic relations. In the deconfined regime, the resulting algebra realizes a mixed 't Hooft anomaly, providing a microscopic route to Abelian crystalline symmetry-enriched topological phases (SETs). These SETs do not mix electric and magnetic sectors, which transform in mutually contragredient representations, thereby preserving their braiding pairing. We illustrate the framework through explicit examples and show that several models present in the literature fit naturally within our construction.

\end{abstract}
	
\maketitle

\tableofcontents
	

\section{Introduction}

\subsection{Motivations}

Generalized symmetries have considerably broadened the framework for characterizing quantum phases of matter. By extending the notion of symmetry beyond conventional global transformations \cite{Batista:2004sc,Nussinov:2006iva,Nussinov:2009zz,Gaiotto:2014kfa,Bhardwaj:2017xup,Chang:2018iay,McGreevy:2022oyu,Gomes:2023ahz,Bhardwaj:2023kri,Shao:2023gho}, they provide a unified language for describing and relating phases with intrinsically nonlocal structures, including topologically ordered states \cite{wen2004quantum,Kitaev:2005hzj,Savary_2017,Wen:2016ddy,Wen:2012hm}. This broader perspective points toward an extension of the Landau paradigm that encompasses higher-form, non-invertible, and subsystem symmetries \cite{Devakul:2018fhz,Iqbal:2021rkn,Bhardwaj:2023fca,Moradi:2022lqp,McGreevy:2022oyu,PhysRevB.106.155150,Chatterjee:2022kxb,Gomes:2023ahz,Seo:2024its,Pace:2025hpb,Chen:2025uno,Ebisu:2026rnu,Giridhar:2025bol}.

An especially interesting class of generalized symmetries is formed by modulated symmetries, whose transformations exhibit a prescribed spatial dependence. Such symmetries arise in a variety of lattice~\cite{Watanabe:2022pgk,PhysRevB.107.195139,Delfino:2023anb,Pace:2022wgl,Delfino:2023rpw,Yoshitome:2025klz,Chen:2025hzm,Pace_2025,Liang:2026upw,Sala:2021jca,Lian:2026rwk,ryb5-8ntp,PhysRevB.110.165109} and continuum systems \cite{Pretko:2016kxt,Pretko:2016lgv,Pretko:2018jbi,Nandkishore:2018sel,PhysRevB.98.125105,Pretko:2018tit,PhysRevLett.120.195301,Seiberg:2020bhn,Seiberg:2020wsg,Seiberg:2020cxy} and can impose unconventional constraints on local operators, leading to structures markedly different from those associated with uniform global transformations. They are also closely connected with subsystem symmetries and gapped fracton phases \cite{PhysRevLett.94.040402,Haah:2011drr,Vijay:2016phm,Slagle:2017wrc,PhysRevResearch.5.043108,Casasola:2023tot,Casasola:2024esm}, some of which may arise through the Higgsing of gapless modulated gauge theories~\cite{PhysRevB.98.035111,Bulmash:2018lid}. A particularly interesting feature of modulated symmetries is their capacity to induce ultraviolet/infrared (UV/IR) mixing, manifested in the sensitivity of their global realization to microscopic lattice scales, for instance through commensurability conditions between the modulation period and the system size \cite{Gorantla:2021bda,Kim:2023qqi,Watanabe:2022pgk}. This intertwining of microscopic spatial structure with the global properties of the system makes modulated symmetries particularly rich from the viewpoint of quantum phases of matter \cite{Ebisu:2023idd,Ebisu:2024eew,Bulmash:2025zua,Pace:2026brj,Ning:2026noc,Saito:2025qrp,Lam:2023xng,Lam:2024smz,Anakru:2026ffh}.

\begin{figure}
	\includegraphics[angle=0,scale=2]{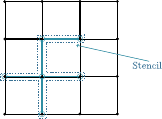}
	\caption{Example of a stencil of size $t=6$. }
    \label{stencil}
\end{figure}

Systematically organizing modulated symmetries requires treating their internal and spatial components on an equal footing~\cite{Lam:2023xng,Pace_2025,Bulmash:2025zua,Pace:2026brj}. This entails establishing the consistency conditions governing their combined algebraic structure and constructing the corresponding bond algebras \cite{PhysRevB.79.214440,PhysRevLett.104.020402,Cobanera:2011wn} of symmetric local Hamiltonians. Gauging the internal symmetry provides a particularly sharp perspective on these structures~\cite{SciPostPhys.19.4.110,Kim:2025pzk,Delfino:2023rpw}. It reorganizes the operator algebra so that the spatial modulation data are inherited by emergent extended operators, directly connecting modulated global symmetries to higher-form symmetries and mixed 't Hooft anomalies~\cite{Kapustin:2014lwa,Kapustin:2014zva}. In two or more spatial dimensions, this framework naturally constructs Abelian crystalline symmetry-enriched topological (SET) phases~\cite{Barkeshli:2014cna,Fidkowski:2016svr}, in which the modulation functions determine how spatial symmetries act on the emergent topological excitations.


\subsection{\label{secB}Overview and Main Results}

In this work, we develop a constructive framework for characterizing discrete Abelian modulated symmetries and their gauging in $d$-dimensional lattice systems with spatial symmetry $G_{\mathrm{space}}$. We consider an internal  symmetry $G_{\mathrm{int}} = \mathbb Z_N^{\times \mathfrak q}$, generated by 
\begin{equation}
U^{(\alpha)} \equiv \prod_{\vec{r}} (X_{\vec{r}})^{f_{\vec{r}}^{(\alpha)}},
\qquad \alpha=1,\ldots,\mathfrak{q},
\end{equation}
where $X_{\vec r}$ are local $\mathbb Z_N$ shift operators that obey $X_{\vec r}^N=1$ for every lattice site $\vec r$.
The spatial dependence is encoded in the modulation functions $f_{\vec{r}}^{(\alpha)}$, whose transformation under the space group $G_{\mathrm{space}}$ closes within the set of $\mathfrak{q}$ generators. In Sec.~\ref{SecGlobal}, we show that this allows the modulation to be formulated as a finite-dimensional representation of the space group. For finite $G_{\mathrm{int}}$, we also show that the nontrivial spatial action necessarily factors through a finite quotient of the space group. Consequently, although the space group $G_{\mathrm{space}}$ itself may be infinite, the modulation pattern is controlled by finite-order transformations and is therefore periodic. Throughout this paper, the period is encoded in the $d-$dimensional integer vector $\vec \tau$ such that $f^{(\alpha)}_{\vec r+\vec \tau} = f_{\vec r}^{(\alpha)}$ for all lattice sites $\vec r$ and function types $\alpha = 1,\ldots, \mathfrak q$.

In Sec.~\ref{sec: symmetry bond algebra}, we provide a systematic construction of the local symmetric bond algebra. For interactions supported on a finite stencil containing $t$ sites, as illustrated in Fig.~\ref{stencil}, the symmetry constraints are encoded in a matrix constructed from the modulation functions, whose kernel yields the symmetry-allowed local terms. When this matrix has maximal row rank and $N$ is prime, the rank--nullity theorem leads to the simple relation between the number of independent symmetries $\mathfrak{q}$, the number of independent local interactions $\chi$, and the size of the stencil $t$,
\begin{equation}
\mathfrak{q}+\chi = t.
\end{equation}
This provides a direct constructive link between the symmetry content of the theory and the local operators from which symmetric Hamiltonians can be built.

A further ingredient of the construction, which we explore in Sec.~\ref{sec: algebra of open operators}, is an algebra of open operators naturally associated with the symmetric bond algebra. Products of local bonds that trivialize along closed sequences define empty operators, while opening these sequences leaves nontrivial contributions at their boundaries. These boundary operators have a nontrivial algebra with truncated versions of the original global symmetry operators. This structure already contains information about possible emergent symmetries  and mixed 't Hooft anomalies in appropriate subalgebras of the ungauged theory.

In Sec.~\ref{sec: gauging procedure} and~\ref{sec: gauge invariang bond algebra}, we show that gauging reorganizes this operator algebra in a particularly transparent way. The original global symmetry operators become trivial in the physical Hilbert space, whereas the empty operators of the ungauged theory are promoted to nontrivial Wilson operators. These Wilson operators generate emergent modulated higher-form quantum symmetries, thereby inheriting the spatial structure of the original modulation. Conversely, restricting the original symmetry operators to finite regions gives rise, after gauging, to nontrivial operators supported on their boundaries. In Sec.~\ref{sec: deconfined phase of gauge theory}, we argue that in the deconfined phase of the gauge theory (pure-gauge limit), the resulting higher-form symmetries possess a nontrivial mutual algebra that realizes a mixed 't Hooft anomaly and provides a natural characterization of the corresponding topologically ordered phase.

In Sec.~\ref{sec: SETs}, we specialize our construction to $d=2$ dimensions, and further determine how $G_{\mathrm{space}}$ acts on the anyonic excitations of the resulting topologically ordered deconfined phase. For orientation-preserving spatial symmetries that leave the electric and magnetic sectors separately invariant, the transformation of the magnetic anyons is inherited from the representation carried by the original modulation functions (Sec.~\ref{sec: transformation of magnetic anyons}). Preservation of the canonical electric--magnetic pairing then fixes the transformation of the electric sector  through the corresponding contragredient representation (Sec.~\ref{sec: transformation of electric anyons}). Thus, the electric and magnetic symmetry actions are not independent. This establishes a direct relation between the modulation data of the ungauged theory and the anyon-permutation structure of the gauged theory, providing a constructive microscopic framework for a broad class of Abelian SET phases (Sec.~\ref{sec: generic anyon transformation}). 

We illustrate the general construction through explicit one- and two-dimensional examples (Secs.~\ref{1D} and \ref{2D}, respectively), including anisotropic and isotropic systems on the square lattice. The formalism also encompasses several models previously studied in the literature, which at first sight arise from rather different modulation patterns and microscopic constructions. Table~\ref{tab:literature} summarizes representative examples and displays their characterization in terms of the common data $(\mathfrak{q},\chi,t)$. In particular, all the examples shown in the table fit naturally into the same kernel construction and obey the relation $\mathfrak{q}+\chi=t$ for the corresponding choice of stencil. The deconfined phase of their gauged counterparts corresponds to SETs enriched by $G_{\mathrm{space}}$ with its modulation data determining the corresponding spatial action on the anyonic excitations. This highlights that the framework developed here not only generates new examples of modulated-symmetry models, but also provides a unified organization of a number of previously known constructions.

\begin{table*}[h]
\centering

\setlength{\tabcolsep}{5pt}
\renewcommand{\arraystretch}{1.4}
\hspace*{-0.8cm}
\begin{tabularx}{\textwidth}{
>{\raggedright\arraybackslash}p{1.3cm}
c c c c c c 
}
\cline{1-7}
\noalign{\vskip\doublerulesep}
\cline{1-7}
Model
& $G_{\text{int}}\rtimes G_{\text{space}}$
& $\mathfrak{q}$
& $\chi$
& $t$
& Modulation
& Period
\\
\cline{1-7}

Ref.~\cite{Watanabe:2022pgk}
& $\mathbb{Z}_N\rtimes T^2$
& $1$
& $2$
& $3$
& $f_{\vec{r}}=a^{i_x+i_y}$
& $(\text{ord}_N(a),\text{ord}_N(a))$
\\
\cline{1-7}

Sec.~\ref{VI}
& $\mathbb{Z}_N^2\rtimes \left(T^2\rtimes C_4\right)$
& $2$
& $2$
& $4$
& $f^{(1)}_{\vec{r}} = \frac{1+(-1)^{i_x+i_y+1}}{2}$,
  $f^{(2)}_{\vec{r}} = \frac{1+(-1)^{i_x+i_y}}{2}$
& (2,2)
\\
\cline{1-7}

Ref.~\cite{Yoshitome:2025klz}
& $\mathbb{Z}_2^2 \rtimes T^2$
& $2$
& $2$
& $4$
& $f^{(1)}_{\vec{r}} = i_x^3 \bmod 3$,
  $f^{(2)}_{\vec{r}} = (i_x+1)^3\bmod 3$
& (3,1)
\\
\cline{1-7}

Ref.~\cite{Pace:2022wgl}
& $\mathbb{Z}_N^3\rtimes \left(T^2\rtimes C_4\right)$
& $3$
& $3$
& $6$
& $f^{(1)}_{\vec{r}} = 1$,
  $f^{(2)}_{\vec{r}} = i_x$,
  $f^{(3)}_{\vec{r}} = i_y$
& $(N,N)$
\\
\cline{1-7}

Ref.~\cite{Delfino:2023anb}
& $\mathbb{Z}_N^4 \rtimes T^2$
& $4$
& $2$
& $6$
& $f^{(1)}_{\vec{r}} = 1$,
  $f^{(2)}_{\vec{r}} = i_x$,
  $f^{(3)}_{\vec{r}} = i_y$,
  $f^{(4)}_{\vec{r}} = i_x i_y$
& $(N,N)$
\\
\cline{1-7}

Ref.~\cite{Delfino:2023rpw}
& $\mathbb{Z}_N^4 \rtimes T^2$
& $4$
& $2$
& $6$
& $f^{(1)}_{\vec{r}} = 1$,
  $f^{(2)}_{\vec{r}} = a^{i_x}$,
  $f^{(3)}_{\vec{r}} = a^{i_y}$,
  $f^{(4)}_{\vec{r}} = a^{i_x+i_y}$
& $(\text{ord}_N(a),\text{ord}_N(a))$
\\
\cline{1-7}
\end{tabularx}

\caption{
Representative realizations of the general modulated-symmetry framework, including models previously discussed in the literature. These are defined on two-dimensional square lattices, where $T^2$ indicates invariance under translations and $C_4$ invariance under $\pi/2$ rotations. The period is the integer vector $\vec{\tau}$ defined in Sec. \ref{secB}.
}
\label{tab:literature}

\end{table*}


\section{Global Modulated Symmetries}\label{SecGlobal}

\subsection{Global Symmetry Group}

We shall discuss the classification of modulated symmetries on $d$-dimensional lattices $\Lambda$ symmetric under space group $G_{\mathrm{space}}$.
We further consider an internal discrete Abelian symmetry group as
\begin{equation}
G_{\text{int}} = \underbrace{\mathbb{Z}_N \times  \mathbb{Z}_N \times \cdots \times  \mathbb{Z}_N}_{\mathfrak{q}~ \text{times}},
\end{equation}
so that its order is $|G_{\text{int}}|=N^{\mathfrak{q}}$. The corresponding global symmetry group is therefore
\begin{equation}\label{full_symmetry}
G= G_{\text{int}} \rtimes G_{\text{space}},
\end{equation} 
where the structure of the semi-direct product will be specified shortly.

The operators that act at each site $\vec{r}\in \Lambda$ are the generalized $\mathbb{Z}_N$ ``clock” $Z$ and ``shift” $X$ Pauli operators, satisfying 
\begin{equation}
	X_{\vec{r}} Z_{\vec{r}'} = \omega^{\delta_{\vec{r},\vec{r}'}} Z_{\vec{r}'} X_{\vec{r}},\label{ZN_algebra}
\end{equation}
with $\omega\equiv e^{\frac{2\pi i}{N}}$. In this way, we can parametrize the unitary symmetry operators of $G_{\text{int}}$ as
\begin{equation}
	U^{(\alpha)} \equiv \prod_{\vec{r}} (X_{\vec{r}})^{f_{\vec{r}}^{(\alpha)}},
    \label{symmetryops}
\end{equation} 
where 
\begin{equation}
f_{\vec{r}}^{(\alpha)}: \Lambda~ \rightarrow~ \mathbb{Z}_N, ~~~\alpha = 1, 2, \ldots, \mathfrak{q}.
\end{equation}
The symmetry generators obey
\begin{eqnarray}
    (U^{(\alpha)})^N = \prod_{\vec r
    } (X_{\vec r})^{N \, f_{\vec r}^{(\alpha)}} = 1,
\end{eqnarray}
i.e., the $\mathbb Z_N$ group relation. The multiplication law is simply
\begin{equation}
U^{(\alpha)} U^{(\beta)} = \prod_{\vec{r}} (X_{\vec{r}})^{f_{\vec{r}}^{(\alpha)}+f_{\vec{r}}^{(\beta)}} \equiv U^{(\gamma)}, 
\end{equation}
with 
\begin{equation}
f_{\vec{r}}^{(\gamma)}=f_{\vec{r}}^{(\alpha)}+f_{\vec{r}}^{(\beta)}.
\end{equation}
If $N$ is prime, which we consider throughout this work, the set $\{f_{\vec{r}}^{(\alpha)}\}$ forms a $\mathfrak{q}$-dimensional vector space over the field $\mathbb Z_N$ under standard function addition and scalar multiplication.

The action of a group element $g \in G_{\text{space}}$ on a site $\vec{r}$ will be denoted by $\vec{r}^g$. This leads to the following transformation of $f_{\vec{r}}^{(\alpha)}$:
\begin{equation}
g: f_{\vec{r}}^{(\alpha)} \rightarrow f_{\vec{r}^g}^{(\alpha)} = \sum_{\beta} A_{\alpha,\beta}^{(g)} f_{\vec{r}}^{(\beta)},
\label{rep}
\end{equation}
where all the matrix elements $A_{\alpha,\beta}^{(g)}$ are defined modulo $N$. This transformation can be understood as follows. Let $U^g_{\mathrm{space}}$ be the unitary operator implementing the spatial symmetry element $g\in G_{\mathrm{space}}$. Acting on the symmetry operator $U^{(\alpha)}$, we obtain
\begin{eqnarray}
U^g_{\mathrm{space}}\,
U^{(\alpha)}\,
U^{g\dagger}_{\mathrm{space}}
&=&
\prod_{\vec r}
U^g_{\mathrm{space}}\,
\left(X_{\vec r}\right)^{f^{(\alpha)}_{\vec r}}\,
U^{g\dagger}_{\mathrm{space}}
\nonumber\\
&=&
\prod_{\vec r}
\left(X_{\vec r}\right)^{
f^{(\alpha)}_{\vec r^{\,g^{-1}}}
}.\label{transf}
\end{eqnarray}
On the other hand, since the set of functions $f^{(\alpha)}$ is assumed to be closed under the action of $G_{\mathrm{space}}$, the transformed function must be expressible as a linear combination of the original ones \cite{Bulmash:2023msp}. Thus, quite generally,
\begin{eqnarray}
U^g_{\mathrm{space}}\,
U^{(\alpha)}\,
U^{g\dagger}_{\mathrm{space}}
=
\prod_{\vec r}
\left(X_{\vec r}\right)^{
\sum_{\beta}
A^{(g^{-1})}_{\alpha\beta}
f^{(\beta)}_{\vec r}
}.
\label{transf1}
\end{eqnarray}
Comparing the two expressions, we find the transformation law in Eq.~\eqref{rep}.

Taking into account the sequence of transformations $\vec{r} \overset{h}{\rightarrow} \vec{r}^h \overset{g}{\rightarrow} \vec{r}^{gh}$, relation \eqref{rep} leads to the homomorphism 
\begin{equation}
A^{(gh)} = A^{(g)} A^{(h)}.
\end{equation}
This implies that such matrices have inverses
\begin{eqnarray}
    A^{(g^{-1})} = (A^{(g)})^{-1},
\end{eqnarray}
showing that $A^{(g)}\in GL(\mathfrak q, \mathbb Z_N)$, i.e., an element of the general linear group.
Thus, the set of matrices $A^{(g)}$ forms a $\mathfrak{q}$-dimensional representation of $G_{\text{space}}$:
\begin{equation}
A: G_{\text{space}}~\rightarrow ~ GL({\mathfrak{q}}, \mathbb{Z}_N),
\end{equation}
i.e., the group of invertible matrices over the field $\mathbb Z_N$. 

The group $GL(\mathfrak q,\mathbb Z_N)$ is finite, with cardinality
\begin{eqnarray}
|GL(\mathfrak q,\mathbb Z_N)|
=
(N^\mathfrak q-1)(N^\mathfrak q-N)\cdots (N^\mathfrak q-N^{\mathfrak q-1}).
\end{eqnarray}
It follows that every element of $GL(\mathfrak q,\mathbb Z_N)$ has finite order. In particular, for any matrix $A^{(g)} \in GL(\mathfrak q,\mathbb Z_N)$, Lagrange's theorem implies that its order must divide $|GL(\mathfrak q,\mathbb Z_N)|$. Therefore, the possible orders are of the form
\begin{eqnarray}\label{order_group_GL}
\tau = N^a \prod_{i=1}^{\mathfrak q} d_i,
\end{eqnarray}
where $a$ is an integer with $0 \le a \le {\mathfrak q(\mathfrak q-1)}/{2}$,
and each $d_i$ is a divisor of $N^i-1$. Consequently, every matrix $A^{(g)}$ has finite order, and this in turn implies that all functions $f_{\vec r}^{(\alpha)}$ are periodic. This is a noteworthy result, since it shows that the modulation structure is controlled not by the full space group $G_{\mathrm{space}}$, but rather by a finite subgroup $H_{\mathrm{space}} \subset G_{\mathrm{space}}$.


 The finiteness of the orders of the matrices $A^{(g)}$ admits a natural group-theoretic interpretation. Indeed, consider the homomorphism
\begin{eqnarray}
\varphi : G_{\mathrm{space}} \to \mathrm{Aut}(G_{\mathrm{int}}), \label{eq: homomorphism varphi}
\end{eqnarray}
which defines the semidirect product $
G_{\mathrm{int}} \rtimes_{\varphi} G_{\mathrm{space}}$
appearing in Eq.~\eqref{full_symmetry}. Since $G_{\mathrm{int}}$ is finite, its automorphism group $\mathrm{Aut}(G_{\mathrm{int}})$ is also finite. It follows that the image of $\varphi$,
\begin{eqnarray}
F \equiv \mathrm{Im}(\varphi) \subseteq \mathrm{Aut}(G_{\mathrm{int}}),
\end{eqnarray}
is a finite group. Thus, although $G_{\mathrm{space}}$ itself may be infinite, its nontrivial action on $G_{\mathrm{int}}$ factors through a finite group.

To make this precise, let
\begin{eqnarray}
K \equiv \ker(\varphi)
= \{g\in G_{\mathrm{space}} \,:\, \varphi(g)=\mathrm{id}_{G_{\mathrm{int}}}\}.
\end{eqnarray}
Then $K$ is a normal subgroup of $G_{\mathrm{space}}$, and by the first isomorphism theorem one has
\begin{eqnarray}
G_{\mathrm{space}}/K \cong F.
\end{eqnarray}
Therefore, the action of $G_{\mathrm{space}}$ on $G_{\mathrm{int}}$ is completely determined by the finite quotient $G_{\mathrm{space}}/K$. 

\subsection{Symmetric Bond-Algebra} \label{sec: symmetry bond algebra}

We consider the set of symmetric operators that are invariant under the symmetry group $G$. Given the parametrization \eqref{symmetryops}, it follows trivially that the shift operators $X_{\vec{r}}$ are automatically invariant. The nontrivial invariant operators centered around site $\vec r\in \Lambda$ can be parametrized as
\begin{eqnarray}
\mathcal{O}_{\vec{r}}^{(a)}\equiv\prod_{\vec s \in \Lambda} (Z_{\vec s})^{\Delta^{(a)}_{\vec r,\vec s}},
\label{local}
\end{eqnarray}
where $\Delta_{\vec r, \vec s}^{(a)}$ are integer weights mod $N$ and 
 $a=1,\ldots,\chi$, where $\chi$ represents the number of distinct symmetry-allowed terms. We shall denote the bond-algebra as
\begin{eqnarray}
    \mathfrak{B}_{\mathrm{matter}} = \left \langle \prod_{\vec s\in \Lambda} (Z_{\vec s})^{\Delta_{\vec r, \vec s}^{(a)}}, ~  X_{\vec r}\right\rangle. \label{bond_algebra}
\end{eqnarray}

To determine the symmetry constraint on $\Delta^{(a)}_{\vec r,\vec s}$, we evaluate the action of $U^{(\alpha)}$ on \eqref{local}.
Using the $\mathbb Z_N$ algebra in \eqref{ZN_algebra}, one finds
\begin{eqnarray}
U^{(\alpha)}
\left(
\prod_{\vec s \in \Lambda} (Z_{\vec s})^{\Delta^{(a)}_{\vec r,\vec s}}
\right)
(U^{(\alpha)})^\dagger
=
\left(
\prod_{\vec s \in \Lambda}
\omega^{\Delta^{(a)}_{\vec r,\vec s} f^{(\alpha)}_{\vec s}}
\right)
\prod_{\vec s \in \Lambda} Z_{\vec s}^{\Delta^{(a)}_{\vec r,\vec s}} .
\label{symmetry_constraint_delta_step}
\end{eqnarray}
Therefore, invariance under $U^{(\alpha)}$ requires the phase factor to be trivial, which is equivalent to the condition
\begin{eqnarray}
\sum_{\vec r' \in \Lambda}
\Delta^{(a)}_{\vec r,\vec r'} f^{(\alpha)}_{\vec r'}
= 0
\qquad
\mathrm{mod}\; N .
\label{symmetry_constraint_delta}
\end{eqnarray}
Hence, the coefficients $\Delta^{(a)}_{\vec r,\vec r'} \in \mathbb Z/N\mathbb Z$ must annihilate the functions $f^{(\alpha)}_{\vec r'}$. This annihilation condition admits extensively many solutions for $\Delta_{\vec r,\vec s}^{(a)}$. Because the corresponding bond-algebra elements are intended to serve as the building blocks for local Hamiltonians, we require additional constraints.

A first such requirement is translation invariance, namely,
\begin{eqnarray}
    \Delta_{\vec r,\vec s} = K_{\vec r-\vec s},
    \label{translation_invariant_delta}
\end{eqnarray}
where we have suppressed the label $a$ to avoid clutter\footnote{Each independent solution obtained by the constructive procedure below gives a distinct matrix $\Delta^{(a)}_{\vec r,\vec s}$, with $a=1,\ldots,\chi$.}.
Thus, $\Delta_{\vec r,\vec s}$ depends only on the relative displacement $\vec r-\vec s$. 
Under periodic boundary conditions, this means that $\Delta_{\vec r, \vec s}$ is a circulant matrix\footnote{An $n\times n$ matrix $C$ is called circulant if it is of the form
\[
    C =
    \begin{pmatrix}
    c_1 & c_2 & \ldots & c_{n-1} & c_n\\
    c_2 & c_3 & \ldots & c_n & c_1\\
    \vdots & \vdots & \ddots & \vdots & \vdots\\
    c_n & c_1 & \ldots & c_{n-2} & c_{n-1}
    \end{pmatrix},
\]
namely, each row is obtained from the preceding one by a cyclic permutation.},
implying that the normalized eigenvectors of $\Delta_{\vec r,\vec s}$ are discrete Fourier modes. 


A further natural requirement is that the interaction terms in Eq.~\eqref{bond_algebra} are local. More precisely, for each reference site $\vec r\in \Lambda$, we require that $\Delta_{\vec r,\vec s}$ have nonvanishing support only on a finite subset $B_t(\vec r)\subset \Lambda$ containing exactly $t$ sites, with $t=\mathcal O(1)$ independent of the system size $|\Lambda|$. Thus,
\begin{eqnarray}
    \Delta_{\vec r,\vec s}=0
    \qquad
    \text{for all}
    \qquad
    \vec s\notin B_t(\vec r),
\end{eqnarray}
and $|B_t(\vec r)|=t,$ for every $\vec r\in\Lambda$. Under this locality assumption, Eq.~\eqref{symmetry_constraint_delta} reduces to
\begin{eqnarray}
    \sum_{\vec s\in B_t(\vec r)}
    K_{\vec r-\vec s}\, f^{(\alpha)}_{\vec s}
    =0
    ~~~
    \mathrm{mod}\, N.
    \label{3430}
\end{eqnarray}
This condition can be written more compactly in matrix form as
\begin{eqnarray}
    F\cdot K = 0
    ~~~
    \mathrm{mod}\, N,
    \label{3431}
\end{eqnarray}
where $F$ is the $\mathfrak q\times t$ matrix with entries $  F_{\alpha,\vec s}\equiv f^{(\alpha)}_{\vec s}$,
for $\vec s\in B_t(\vec r)$, and $K$ is the $t\times 1$ column vector whose components are $K_{\vec r-\vec s}$. This provides a constructive characterization of the allowed solutions $\Delta_{\vec r,\vec s}^{(a)}$: once the matrix $F$ is assembled from the modulation functions $f^{(\alpha)}_{\vec s}$, the admissible local interactions are obtained by determining its kernel over $\mathbb Z/N\mathbb Z$.


This matrix formulation is especially convenient, as it makes explicit the relation between the number of admissible solutions $\chi$ and the number of independent modulation functions $\mathfrak q$. In fact, the number of solutions is given by
\begin{eqnarray}
    \chi = \dim \ker(F),
\end{eqnarray}
that is, by the number of linearly independent vectors $K$ satisfying $F\cdot K=0$ over $\mathbb Z_N$.

Furthermore, by assumption the $\mathfrak q$ functions $f^{(\alpha)}$ are linearly independent. Hence, for sufficiently large support size $t$, the matrix $F$ attains maximal row rank,
\begin{eqnarray}
    \mathrm{rank}(F)=\mathfrak q.
\end{eqnarray}
Applying the rank--nullity theorem then yields
\begin{eqnarray}
\label{ranknullity}
    \mathfrak q+\chi=t,
\end{eqnarray}
which constrains the number of independent functions $\mathfrak q$, the number of bond-algebra elements $\chi$, and the support size $t$ of the local interaction terms.


A few caveats are worth emphasizing when determining the support size $t$ of the neighborhood $B_t(\vec r)$. On the one hand, $t$ must be large enough so that the restricted functions $f^{(\alpha)}_{\vec s}$, for $\vec s\in B_t(\vec r)$, remain linearly independent, as this is a crucial assumption in the argument above. On the other hand, if $t$ is taken unnecessarily large, one may encounter additional solutions for $K$ that are not fundamentally new, but are instead generated by translations or linear combinations of solutions already obtained at smaller support size. A detailed discussion of how the optimal value of $t$ is fixed in concrete examples is presented in Secs. \ref{1D} and \ref{2D}.


Note that the vectors $K$ belong not only to the kernel of $F$, but also to the kernel of the transformed matrix
\begin{eqnarray}
    \widetilde F^{(g)} = A^{(g)}\cdot F ,
\end{eqnarray}
for every $g\in G_{\mathrm{space}}$. This follows from the assumption that the set of $\mathfrak q$ modulation functions is closed under the action of $G_{\mathrm{space}}$. Consequently, the space of solutions is itself closed under spatial symmetries: whenever a given interaction bond is allowed, all of its symmetry-related images must also be allowed.
The transformation properties of the local operators in Eq.~\eqref{local} can be determined in a way analogous to the derivation of Eq.~\eqref{transf}. Acting with the spatial symmetry operator $U^g_{\mathrm{space}}$ on $\mathcal{O}_{\vec r}^{(a)}$ of \eqref{local}, we find
\begin{eqnarray}
U^g_{\mathrm{space}}\,
\mathcal O_{\vec r}^{(a)}\,
U^{g\dagger}_{\mathrm{space}}
&=&
U^g_{\mathrm{space}}
\left[
\prod_{\vec s\in\Lambda}
\left(Z_{\vec s}\right)^{\Delta^{(a)}_{\vec r,\vec s}}
\right]
U^{g\dagger}_{\mathrm{space}}
\nonumber\\
&=&
\prod_{\vec s\in\Lambda}
\left(Z_{\vec s^{\,g}}\right)^{\Delta^{(a)}_{\vec r,\vec s}}
\nonumber\\
&=&
\prod_{\vec s\in\Lambda}
\left(Z_{\vec s}\right)^{\Delta^{(a)}_{\vec r,\vec s^{\,g^{-1}}}} .
\end{eqnarray}
On the other hand, since the transformed operator must again belong to the same solution space, it can be expanded in the basis of allowed bonds as
\begin{eqnarray}
U^g_{\mathrm{space}}\,
\mathcal O_{\vec r}^{(a)}\,
U^{g\dagger}_{\mathrm{space}}
=
\prod_{\vec s\in\Lambda}
\left(Z_{\vec s}\right)^{
\sum_b M_{ab}^{(g^{-1})}
\Delta^{(b)}_{\vec r^{\,g},\vec s}
}.
\end{eqnarray}
Comparing the exponents of $Z_{\vec s}$ in the two expressions gives
\begin{eqnarray}
\Delta^{(a)}_{\vec r,\vec s^{\,g^{-1}}}
=
\sum_b
M_{ab}^{(g^{-1})}
\Delta^{(b)}_{\vec r^{\,g},\vec s}.
\end{eqnarray}
Equivalently, after relabeling the spatial arguments, this can be written as
\begin{eqnarray}
\Delta^{(a)}_{\vec r^{\,g},\vec s^{\,g}}
=
\sum_{b=1}^{\chi}
M^{(g)}_{ab}\,
\Delta^{(b)}_{\vec r,\vec s}.
\label{deltatransf}
\end{eqnarray}
This equation defines the $\chi\times\chi$ matrix $M^{(g)}$, which describes the induced action of $G_{\mathrm{space}}$ on the solutions $\Delta_{\vec r, \vec s}$. Similar to the case of the matrices $A^{(g)}$ introduced in Eq.~\eqref{rep}, the matrices $M^{(g)}$ furnish a $\chi$-dimensional representation of $G_{\mathrm{space}}$.


We assume throughout that each spatial symmetry element $g \in G_{\mathrm{space}}$ permutes the interaction types, up to an optional reversal of the corresponding generators. This action is encoded by a permutation $\pi_g$ of $\{1,\ldots,\chi\}$ and signs $\epsilon_g(a)\in\{+1,-1\}$. In particular, $\epsilon_g(a)=-1$ means that the interaction associated with
$a$ is mapped to the inverse of the interaction associated with
$\pi_g(a)$. Equivalently, the interaction types furnish a monomial
representation
\begin{equation}
M^{(g)}_{ab}
    =
    \epsilon_g(b)\,
    \delta_{a,\pi_g(b)}.
    \label{monomial}
\end{equation}
The group-composition law is
\begin{equation}
M^{(gh)}=M^{(g)}M^{(h)},
\end{equation}
or, equivalently,
\begin{equation}
    \pi_{gh}=\pi_g\circ\pi_h,
    \qquad
    \epsilon_{gh}(a)
    =
    \epsilon_h(a)\,
    \epsilon_g\!\left(\pi_h(a)\right).
\end{equation}
In particular, $\pi_{g^{-1}}=\pi_g^{-1}$, together with the corresponding
inverse relation for the signs.
With this, the general transformation \eqref{deltatransf} reduces to 
\begin{equation}
    \Delta^{(\pi_g(a))}_{\vec r^{\,g},\vec s^{\,g}}
    =
    \epsilon_g(a)\,
    \Delta^{(a)}_{\vec r,\vec s}.
    \label{eq:Delta_covariance}
\end{equation}

Although spatial transformations could, in general, mix interaction types via an arbitrary matrix $M \in \mathrm{GL}(\chi, \mathbb{Z}_N)$, the approach adopted here introduces a substantial simplification. This framework captures a broad class of systems, including all literature examples considered in this work.


\subsection{Algebra of Open Operators} \label{sec: algebra of open operators}

We now discuss an algebraic structure that is implicit in the above construction. As we will see, this structure provides a useful way to characterize the symmetry pattern both before and after gauging. In the ungauged theory, it captures nontrivial commutation relations among certain operators and reveals possible emergent symmetries associated with suitable subalgebras of the bond algebra \eqref{bond_algebra}. After gauging, the same algebraic relations are reorganized into the operator structure of the gauged theory, where they are naturally connected to Wilson operators and emergent higher-form quantum symmetries.

The first step in characterizing this structure is to identify the \textit{empty operators} associated with the bond algebra in Eq.~\eqref{bond_algebra}. These operators are products of local matter bonds whose net contribution is trivial, i.e., their product reduces to the identity operator \cite{Chatterjee:2022kxb}. Since the interaction terms
\begin{eqnarray}
\mathcal{O}_{\vec{r}}^{(a)}\equiv \prod_{\vec s\in \Lambda}
(Z_{\vec s})^{\Delta^{(a)}_{\vec r,\vec s}}=
\prod_{\vec s\in B_t^{(a)}(\vec r)}
(Z_{\vec s})^{\Delta^{(a)}_{\vec r,\vec s}}
\end{eqnarray}
are local operators supported on the finite stencil $B_t^{(a)}(\vec r)$, they can be combined in such a way that their product trivializes to the identity, resulting in an empty operator. Concretely, an empty operator is obtained by multiplying such bonds along a closed path of neighboring stencils, defined by a sequence
\begin{eqnarray}
    \gamma
    =
    \big\{
    (\vec r_1,a_1),
    (\vec r_2,a_2),
    \ldots,
    (\vec r_n,a_n)
    \big\},
    \label{gamma1}
\end{eqnarray}
where $a_i$ labels the interaction type of the operator $\mathcal{O}_{\vec{r}_i}^{(a_i)}$ centered at site $\vec{r}_i \in \Lambda$. The sequence is
constructed so that neighboring stencils overlap
\begin{eqnarray}
    \vec r_{i+1} \in B_t^{(a_{i})}(\vec r_{i}),
\end{eqnarray}
with the additional closing condition
\begin{eqnarray}
    \vec r_{n+1} \in B_t^{(a_1)}(\vec r_1).
    \label{pbc}
\end{eqnarray}
More precisely, we need to include integer-valued functions $g^{(a_i,\bar{\alpha})}_{\vec r_i}\in\mathbb Z/N\mathbb Z$ weights along the path defined by the sequence, namely, 
\begin{equation}
W^{(\bar{\alpha})}(\gamma)\equiv\prod_{i=1}^n(\mathcal{O}_{\vec r_i}^{(a_i)})^{g_{\vec{r}_i}^{(a_i,\bar{\alpha})}}.
\label{345}
\end{equation}
The condition that the product in \eqref{345} trivializes locally is
\begin{eqnarray}
    \sum_{i=1}^n
     g_{\vec r_i}^{(a_i,\bar{\alpha})}
    \Delta^{(a_i)}_{\vec r_i,\vec s}
    =
    0
    \qquad
    \mathrm{mod}\, N,
    \qquad
    \forall\, \vec s\in\Lambda .
    \label{empty_operator_condition}
\end{eqnarray}
Note that, at this stage, we have not yet specified the nature of the index $\bar{\alpha}$. We will clarify its meaning shortly.


It is natural to ask why such trivial operators can be useful. There are two main reasons. First, in the ungauged theory, they give rise to nontrivial operators when the string $\gamma$ is taken to be open; these {\it open} operators play a crucial role in our construction, as discussed below. Second, in the gauged theory, the empty operators are directly tied to the emergent higher-form symmetries. We shall return to this point later.

Both the symmetry operators $U^{(\alpha)}$ and empty operators $W^{(\bar\alpha)}(\gamma)$ are defined on regions with no boundaries. The symmetry operators $U^{(\alpha)}$ have support on the entire periodic lattice, while $W^{(\bar\alpha)}(\gamma)$ are defined along closed strings $\gamma$. We now consider their open counterparts, which we refer to as \textit{open operators}. These are operators with boundaries or endpoints. 

In particular, the open counterpart of an empty operator is no longer trivial. Indeed, instead of imposing the closure condition in Eq.~\eqref{pbc}, one allows the corresponding sequence of bonds to terminate at nontrivial endpoints. As a result, the product does not trivialize globally, but leaves behind localized endpoint contributions,
\begin{eqnarray}
W^{(\bar\alpha)}(\gamma_{\text{open}})&=& \prod_{\vec{s}_1\in \partial\gamma(\vec{r}_1)}(Z_{\vec{s}_1})^{\sum_{i=1}^n g^{(a_i,\bar\alpha)}_{\vec{r}_i}\Delta_{\vec{r}_i,\vec{s}_1}^{(a_i)} } \prod_{\vec{s}_n\in \partial\gamma(\vec{r}_n)}(Z_{\vec{s}_n})^{\sum_{i=1}^n  g^{(a_i,\bar\alpha)}_{\vec{r}_i}\Delta_{\vec{r}_i,\vec{s}_n}^{(a_i)} }\nonumber\\
&\equiv & W_{\partial}^{(\bar\alpha)}(\vec{r}_1) W_{\partial}^{(\bar\alpha)}(\vec{r}_n), 
\end{eqnarray}
where $\partial\gamma(\vec{r}_1)$ and $\partial\gamma(\vec{r}_n)$ are small regions around the endpoints. 
The notation in the last line emphasizes that the empty operator becomes a nontrivial boundary operator supported at the endpoints of $\gamma_{\text{open}}$. We can also define an open version of the symmetry operators $U^{(\alpha)}$ by restricting their support to a finite $d$-dimensional region $\Gamma\subset\Lambda$:
\begin{equation}
U^{(\alpha)}(\Gamma)=
\prod_{\vec r\in\Gamma}
\left(X_{\vec r}\right)^{f_{\vec r}^{(\alpha)}}.
\end{equation}

\begin{figure}
	\includegraphics[angle=0,scale=0.7]{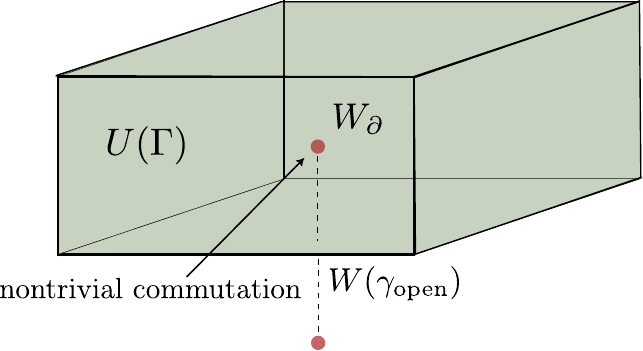}
	\caption{Nontrivial commutation relation between the open operators $U(\Gamma)$ and $W(\gamma_{\text{open}})$. The solid green rectangular prism denotes the $d$-dimensional support  $\Gamma$ of the truncated symmetry operator, while the dotted line represents the empty segment of the open operator $W(\gamma_{\mathrm{open}})$. The red dots indicate its nontrivial endpoints $W_{\partial}$.}
    \label{mix}
\end{figure}

The above two types of open operators can intersect nontrivially when one endpoint of $\gamma_{\mathrm{open}}$ lies inside the region $\Gamma$, while the other lies outside. This intersection is illustrated in Fig.~\ref{mix}. Taking $\vec r_1\in\Gamma$ and $\vec r_n\notin\Gamma$, the operators satisfy the relation
\begin{equation}
U^{(\alpha)}(\Gamma) W_{\partial}^{(\bar\alpha)}(\vec{r}_1) = \omega^{\sum_{\vec{s}\in \partial\gamma(\vec{r}_1)}\sum_{i=1}^n g_{\vec{r}_i}^{(a_i,\bar\alpha)}\Delta^{(a_i)}_{\vec{r}_i,\vec{s}}f_{\vec{s}}^{(\alpha)}} W_{\partial}^{(\bar\alpha)}(\vec{r}_1) U^{(\alpha)}(\Gamma).
\label{mixed}
\end{equation}
This shows that the endpoint operators $W_{\partial}^{(\bar\alpha)}(\vec{r}_1)$ play the role of order parameters for the symmetries generated by $U^{(\alpha)}$.

To characterize these endpoint operators, consider the nontrivial commutation relation localized around $\vec{r}_1$, which is determined by
\begin{equation}
\sum_{\vec{s} \in (\Gamma \cap\partial\gamma(\vec{r}_1))}\,\sum_{\vec{r}_i\in (\gamma \cap B_t(\partial\gamma(\vec{r}_1)\cap \Gamma))}
g_{\vec{r}_i}^{(a_i,\bar\alpha)}
\Delta^{(a_i)}_{\vec{r}_i,\vec{s}}
f_{\vec{s}}^{(\alpha)}.
\end{equation}
For each fixed $\bar\alpha$, this defines a set of $\mathfrak{q}$ local conditions, one for each symmetry label $\alpha$, involving only the coefficients $g_{\vec{r}_i}^{(a_i,\bar\alpha)}$ associated with a local region around $\vec{s}_1$. We therefore choose a set of local coefficients satisfying
\begin{equation}
\sum_{\vec{s} \in (\Gamma \cap\partial\gamma(\vec{r}_1))}\,\sum_{\vec{r}_i\in (\gamma \cap B_t(\partial\gamma(\vec{r}_1)\cap \Gamma))}
g_{\vec{r}_i}^{(a_i,\bar\alpha)}
\Delta^{(a_i)}_{\vec{r}_i,\vec{s}}
f_{\vec{s}}^{(\alpha)}
=\mathcal{I}(\partial\gamma, \Gamma)
\delta_{\alpha,\bar\alpha},
\label{crossing_global}
\end{equation}
with $\bar\alpha=1,\ldots,\mathfrak{q}$ and $\mathcal{I}(\partial\gamma, \Gamma)$ characterizing the intersection. In this way, the label $\bar\alpha$ distinguishes the independent local endpoint operators according to their commutation relations with the symmetry generators $U^{(\alpha)}$. The local order parameters $W_{\partial}^{(\bar\alpha)}$ can thus be chosen in one-to-one correspondence with the symmetry operators $U^{(\alpha)}$. Once the coefficients at one endpoint are fixed, the coefficients $g_{\vec{r}_i}^{(a_i,\bar\alpha)}$ along the entire open string $\gamma_{\mathrm{open}}$ are determined by the recurrence relations in Eq.~\eqref{empty_operator_condition}. In other words, given a local endpoint operator $W_{\partial}^{(\bar\alpha)}$ and a fixed open string $\gamma_{\mathrm{open}}$, we obtain the corresponding extended operator $W^{(\bar\alpha)}(\gamma_{\mathrm{open}})$.


An immediate consequence of the algebra \eqref{mixed} is the following. Consider the subalgebra of Eq.~\eqref{bond_algebra} generated only by the bond operators
\begin{eqnarray}
\mathfrak{B}_{\text{sub}}\equiv \left\langle
\prod_{\vec s\in\Lambda}
\left(Z_{\vec s}\right)^{\Delta^{(a)}_{\vec r,\vec s}} \right\rangle.
\end{eqnarray}
In a phase whose representative Hamiltonian is built entirely from elements of this subalgebra, the boundary operators $W_{\partial}^{(\bar\alpha)}(\vec{s})$ commute with the Hamiltonian and therefore become additional $d$-form symmetries.
In this gapped phase, the algebraic relation in Eq.~\eqref{mixed} can be interpreted as a mixed 't Hooft anomaly between the original $0$-form symmetry and an emergent $d$-form symmetry. The presence of this anomaly constrains the infrared behavior of the theory and, in the present setting, enforces an ordinary spontaneous symmetry breaking phase, which can be characterized by a local order parameter associated with the $d$-form symmetry. The resulting symmetry pattern is
\begin{equation}
G
=
\left(
G_{\mathrm{int}}^{(0)}
\rtimes
G_{\mathrm{int}}^{(d)}
\right)
\rtimes
G_{\mathrm{space}}.
\label{kkjjhh}
\end{equation}


Once the symmetries generated by $U^{(\alpha)}$ are gauged, the operators $U^{(\alpha)}$ act trivially on the physical Hilbert space and hence become empty operators of the resulting theory. Conversely, the empty operators of the original theory are promoted to nontrivial operators. Thus, gauging effectively interchanges the roles of the original symmetry operators and the empty operators. This gives rise to a distinct realization of the nontrivial commutation relation \eqref{mixed}  (Fig.~\ref{mix}), leading to $G_{\mathrm{space}}$- enriched topologically ordered deconfined phases, as we discuss in the following section.

\section{Gauging Modulated Symmetries}\label{SecGauging}

We now turn to the problem of gauging such modulated symmetries. Let $\mathfrak T$ be a theory with modulated symmetry $G = G_{\mathrm{int}}\rtimes G_{\mathrm{space}}$, in Eq.~\eqref{full_symmetry}. Gauging a modulated symmetry, for us,  means gauging the internal subgroup $G_{\mathrm{int}}$, a normal subgroup of $G$. In general, gauging $G_{\mathrm{int}}$ can be understood as a procedure that maps the original theory to a new one,
\begin{eqnarray}
    \mathfrak T \longmapsto \mathfrak T/G_{\mathrm{int}},
\end{eqnarray}
a modulated gauge theory. This operation typically produces an emergent dual symmetry $G^\vee$ of the gauged theory $\mathfrak T/G_{\text{int}}$ referred to as quantum symmetry. Because $G_{\mathrm{int}}$ is Abelian, the associated quantum symmetry is invertible: it is given by the dual group of characters of $G_{\mathrm{int}}$, which is again a finite Abelian group ($\mathrm{Rep}(G_{\mathrm{int}})\cong G_{\mathrm{int}}$).

A key point, however, is that gauging changes not only the symmetry group acting on the theory, but also the form degree of the emergent symmetry. In $d$ spatial dimensions, gauging a $p$-form symmetry produces a dual quantum symmetry of degree $d-p-1$. Since the modulated internal symmetries considered here are $0$-form symmetries, the gauged theory is expected to exhibit emergent $(d-1)$-form quantum symmetries. Thus, even though the gauged subgroup is an ordinary internal symmetry, the resulting quantum symmetry $G^\vee$ is naturally higher-form in spatial dimensions $d>1$.


\subsection{Gauging Procedure}\label{sec: gauging procedure}

The gauging procedure can be formulated as a two-step process. One first extends the Hilbert space by introducing new gauge degrees of freedom, and then imposes a Gauss-law constraint selecting the gauge-invariant physical subspace. In the case of the Abelian discrete internal symmetries $G_{\mathrm{int}}$ studied here, the natural gauge variables are $\mathbb Z_N$ degrees of freedom placed on a dual lattice $\Lambda^\vee$. 


Let us consider the symmetric bond algebra in Eq.~\eqref{bond_algebra}. The matrices $\Delta^{(a)}_{\vec r,\vec s}$ specify the 
interaction terms and have nonvanishing support only on the corresponding stencil $B^{(a)}_t(\vec r)$. The dual lattice $\Lambda^\vee$ is then defined by assigning a dual site to the geometric center of each such stencil. Depending on the geometry of the interaction, these dual sites may be naturally associated with vertices, edges, faces, volumes, or higher-dimensional cells of the original lattice $\Lambda$.

Since there are $\chi$ independent bond-algebra generators associated with each reference site $\vec r\in\Lambda$, the dual lattice contains $\chi$ dual sites for each site of the original lattice. Equivalently, the dual lattice decomposes into a disjoint union of $\chi$ sublattices,
\begin{eqnarray}
    \Lambda^\vee
    =
    \Lambda^\vee_1
    \sqcup
    \Lambda^\vee_2
    \sqcup
    \cdots
    \sqcup
    \Lambda^\vee_\chi .
\end{eqnarray}
Each sublattice $\Lambda^\vee_a$ corresponds to one family of stencils, or equivalently to one independent generator of the bond algebra. This structure defines, for each $a=1,\ldots,\chi$, a one-to-one map
\begin{eqnarray}
\mu_a:\Lambda &\longrightarrow& \Lambda^\vee_a ,
\nonumber\\
\vec r &\longmapsto& \mu_a(\vec r)\equiv \vec{r}_a,
\label{0090}
\end{eqnarray}
which assigns to each site $\vec r\in\Lambda$ a dual site $\vec{r}_a$ associated with the stencil $B^{(a)}_t(\vec r)$. Accordingly, the gauge degrees of freedom living in the dual lattice will be denoted by 
\begin{equation}
\mathcal X_{\vec r_a}^{(a)} ~~~\text{and}~~~ \mathcal Z_{\vec r_a}^{(a)},
\end{equation}
where it is convenient to explicitly keep the superscript $(a)$ to indicate the gauge field flavor.
They obey the algebra
\begin{eqnarray}
\mathcal X_{\vec r_a}^{(a)}\mathcal Z_{\vec r_b}^{(b)}
=
\omega^{\delta_{ab}\delta_{\vec{r}_a,\vec{r}_b}}\,
\mathcal Z_{\vec r_b}^{(b)}\mathcal X_{\vec r_a}^{(a)}.
\end{eqnarray}
Their corresponding transformations under $G_{\text{space}}$, according to the monomial representation \eqref{monomial}, are given by
\begin{equation}
 U_{\text{space}}^g   \mathcal{X}_{\vec r_a}^{(a)} U_{\text{space}}^{q\dagger}=\left(\mathcal{X}_{\vec r_a^g}^{(\pi_g(a))}\right)^{\epsilon_g(a)}~~~\text{and}~~~U_{\text{space}}^g   \mathcal{Z}_{\vec r_a}^{(a)} U_{\text{space}}^{q\dagger}=\left(\mathcal{Z}_{\vec r_a^g}^{(\pi_g(a))}\right)^{\epsilon_g(a)}.
 \label{gaugetransf}
\end{equation}

We denote the inverse map by
\begin{eqnarray}
\nu_a:\Lambda^\vee_a &\longrightarrow& \Lambda ,
\nonumber\\
\vec{r}_a &\longmapsto& \nu_a(\vec{r}_a)\equiv\vec{r}.
\end{eqnarray}
Thus, $\nu_a\circ \mu_a = 1_{\Lambda},
    $ and $
    \mu_a\circ \nu_a = 1_{\Lambda^\vee_a}$
for each $a=1,\ldots,\chi$.



The enlarged Hilbert space contains both the original matter degrees of freedom and the newly introduced gauge degrees of freedom:
\begin{eqnarray}
\mathcal H_{\mathrm{tot}}
=
\mathcal H_{\mathrm{matter}}
\otimes
\mathcal H_{\mathrm{gauge}},
\end{eqnarray}
where
\begin{eqnarray}
\mathcal H_{\mathrm{matter}}
=
\bigotimes_{\vec r\in \Lambda}
\mathbb C^N_{\vec r}~~~\text{and}~~~
\mathcal H_{\mathrm{gauge}}
=
\bigotimes_{a=1}^{\chi}
\bigotimes_{\vec r_a\in \Lambda^\vee_a}
\mathbb C^N_{\vec r_a}.
\end{eqnarray}
Thus, the gauged theory may be viewed as a theory defined on the enlarged lattice $\Lambda \sqcup \Lambda^\vee$,
where $\mathbb Z_N$ matter degrees of freedom live on the original lattice $\Lambda$, while $\mathbb Z_N$ gauge degrees of freedom live on the dual lattice $\Lambda^\vee$. The physical Hilbert space, however, is a subset $\mathcal H_{\mathrm{phys}}\subset \mathcal H_{\mathrm{tot}}$ and is selected by imposing the Gauss-law constraints
\begin{eqnarray}
G_{\vec r}
=
X_{\vec r}^\dagger
\prod_{a=1}^{\chi}
\prod_{\vec s_a\in\Lambda^{\vee}}
\left(
\mathcal X_{\vec s_a}^{(a)}
\right)^{
\Delta^{(a)}_{\vec s,\vec r}
},
\qquad
\vec r\in\Lambda ~~~\text{and}~~~ \vec{s}=\nu_a(\vec s_a).
\label{Gauss_law}
\end{eqnarray}
More explicitly, we define
\begin{eqnarray}
\mathcal H_{\mathrm{phys}}
=
\left\{
|\psi\rangle \in \mathcal H_{\mathrm{tot}}
\;\big|\;
G_{\vec r}|\psi\rangle = |\psi\rangle
\quad
\forall\, \vec r\in\Lambda
\right\},
\end{eqnarray}
namely, the subspace invariant under the Gauss law.


\subsection{Gauge-Invariant Bond Algebra}
\label{sec: gauge invariang bond algebra}

We are now ready to describe the class of theories compatible with the gauge structure introduced above. The most general gauge-invariant bond algebra is
\begin{eqnarray}
\mathfrak B_{\mathrm{gauge\text{-}inv}}
=
\left\langle
\mathcal Z_{\vec r_a}^{(a)}
\prod_{\vec s\in\Lambda}
\left(Z_{\vec s}\right)^{\Delta^{(a)}_{\vec r,\vec s}},
~
X_{\vec r},
~
\mathcal X_{\vec r_a}^{(a)}
\right\rangle,
\label{coupled_bond_algebra}
\end{eqnarray}
which includes pure-gauge operators $\mathcal X_{\vec r_a}$ that generate quantum fluctuations of the gauge degrees of freedom.

Every generator of $\mathfrak B_{\mathrm{gauge-inv}}$ commutes with $G_{\vec r}$. Both $X_{\vec r}$ and $\mathcal X_{\vec r_a}^{(a)}$ commute with the Gauss law trivially. The only nontrivial check involves the minimally coupled bond. Using the $\mathbb Z_N$ algebra for both matter and gauge degrees of freedom, one finds
\begin{align}
G_{\vec r}\,\left(
\mathcal Z_{\vec s_b}^{(b)}
\prod_{\vec s'\in\Lambda}
Z_{\vec s'}^{\Delta^{(b)}_{\vec s,\vec s'}}
\right)
&=
\left(
X_{\vec r}^\dagger
\prod_{a=1}^{\chi}
\prod_{\vec r_a'\in\Lambda^{\vee}}
\left(\mathcal X_{\vec r_a'}^{(a)}\right)^{\Delta^{(a)}_{\vec r',\vec r}}
\right)
\left(
\mathcal Z_{\vec s_b}^{(b)}
\prod_{\vec s'\in\Lambda}
Z_{\vec s'}^{\Delta^{(b)}_{\vec s,\vec s'}}
\right)
\nonumber\\
&=
\omega^{-\Delta^{(b)}_{\vec s,\vec r}}\,
\omega^{+\Delta^{(b)}_{\vec s,\vec r}}\,
\left(
\mathcal Z_{\vec s_b}^{(b)}
\prod_{\vec s'\in\Lambda}
Z_{\vec s'}^{\Delta^{(b)}_{\vec s,\vec s'}}
\right)
\left(
X_{\vec r}^\dagger
\prod_{a=1}^{\chi}
\prod_{\vec r_a'\in\Lambda^{\vee}}
\left(\mathcal X_{\vec r_a'}^{(a)}\right)^{\Delta^{(a)}_{\vec r',\vec r}}
\right)
\nonumber\\
&=
\left(
\mathcal Z_{\vec s_b}^{(b)}
\prod_{\vec s'\in\Lambda}
Z_{\vec s'}^{\Delta^{(b)}_{\vec s,\vec s'}}
\right)\,G_{\vec r},
\qquad
\forall\,\vec r,\vec s\in\Lambda,
\end{align}
ensuring that the minimally coupled bond is gauge invariant.


It is convenient to note that within $\mathcal H_{\mathrm{phys}}$, the Gauss-law condition implies the operator identity
\begin{eqnarray}
X_{\vec r}
=
\prod_{a=1}^{\chi}
\prod_{\vec s_a\in\Lambda^{\vee}}
\left(
\mathcal X_{\vec s_a}^{(a)}
\right)^{
\Delta^{(a)}_{\vec s,\vec r}
},
\qquad
\forall\vec r\in\Lambda .
\label{eq:matter_gauge_constraint}
\end{eqnarray}


We now turn to the fate of the symmetries in the gauged theory. First, note that the symmetries $G_{\mathrm{int}}$ of the original theories $\mathfrak T$ are now fully gauged $
U^{(\alpha)} \big|_{\mathcal{H}_{\mathrm{phys}}} = 1$ for $\alpha=1, \ldots, \mathfrak q$.
To see this explicitly, note that inside $\mathcal H_{\mathrm{phys}}$ we can use \eqref{eq:matter_gauge_constraint} and thus
\begin{eqnarray}
    U^{(\alpha)}\big|_{\mathcal H_{\mathrm{phys}}} & = & \prod_{\vec r \in \Lambda} X_{\vec r}^{f^{(\alpha)}_{\vec r}} = \prod_{\vec r\in \Lambda}  \left[\prod_{a=1}^{\chi}
\prod_{\vec s_a\in\Lambda^{\vee}}
\left(
\mathcal X_{\vec s_a}^{(a)}
\right)^{
\Delta^{(a)}_{\vec s,\vec r}f_{\vec{r}}^{(\alpha)}
}\right]\nonumber\\
    &=& \prod_{a=1}^\chi
\prod_{\vec s_a\in \Lambda^\vee_a}
(\mathcal X_{\vec s_a}^{(a)})^{\sum_{\vec{r}\in \Lambda}
    \Delta^{(a)}_{\vec s,\,\vec r} \, f_{\vec r}^{(\alpha)}}=1,\label{trivialization_gauge}
\end{eqnarray}
where we used that $f^{(\alpha)}\in \operatorname{Ker}(\Delta)$, according to \eqref{symmetry_constraint_delta}. This is the statement that any theory in $\mathfrak B_{\mathrm{gauge-inv}}$ no longer has $G_{\mathrm{int}}$ symmetry. In other words, the operators $U^{(\alpha)}$ become empty operators after gauging.



On the other hand, the gauged theories $\mathfrak T/G_{\mathrm{int}}$ generated by $\mathfrak B_{\mathrm{gauge-inv}}$ possess emergent higher-form quantum symmetries for suitable subalgebras of $\mathfrak B_{\mathrm{gauge-inv}}$. To see this, let us consider the minimally coupled operator in Eq. \eqref{coupled_bond_algebra}, 
\begin{eqnarray}
\tilde{\mathcal{O}}^{(a)}_{\vec r_a}
\equiv
\mathcal Z_{\vec r_a}^{(a)}
\prod_{\vec s\in\Lambda}
Z_{\vec s}^{\Delta^{(a)}_{\vec r,\vec s}}
\in
\mathfrak B_{\mathrm{gauge-inv}},
\qquad
\vec r_a=\mu_a(\vec r).
\end{eqnarray}
Then, for each dual modulation function $g^{(\bar\alpha)}_{\vec r}$, with $\bar\alpha=1,\ldots,\mathfrak q$, and for any closed string $\gamma$
of the type \eqref{gamma1}, we can construct the gauge-coupled counterpart of the operator \eqref{345},
\begin{eqnarray}
W^{(\bar\alpha)}(\gamma)
&=&
\prod_{i=1}^n
\left(
\tilde{\mathcal{O}}^{(a_i)}_{\vec r_{a_i}}
\right)^{ g^{(a_i, \bar\alpha)}_{\vec r_i}}
\nonumber\\
&=&
\prod_{i=1}^n
\left(
\mathcal Z_{\vec r_{a_{i}}}^{(a_i)}
\right)^{g^{(a_i,\bar\alpha)}_{\vec r_i}},\qquad
\vec r_{a_i}=\mu_{a_i}(\vec r_i).
\label{Wilson}
\end{eqnarray}
In the second line, the matter contributions cancel as a consequence of Eq.~\eqref{empty_operator_condition}. Thus, $W^{(\bar\alpha)}(\gamma)$ depends only on the gauge degrees of freedom supported along the closed string $\gamma$. Note that $W^{(\bar\alpha)}(\gamma)$ are no longer empty operators in the gauged theory.
Furthermore, they commute with all elements of the subalgebra $\mathfrak B_{\mathrm{iso}}\subset \mathfrak B_{\mathrm{gauge-inv}}$,
\begin{eqnarray}
\mathfrak B_{\mathrm{iso}}
=
\left\langle
\mathcal Z_{\vec r_a}^{(a)}
\prod_{\vec s\in\Lambda}
\left(Z_{\vec s}\right)^{\Delta^{(a)}_{\vec r,\vec s}},
~
X_{\vec r}
\right\rangle,
\label{iso}
\end{eqnarray}
which is isomorphic to the original bond algebra in Eq.~\eqref{bond_algebra} \cite{Moradi:2023dan}.

The operators in Eq. \eqref{Wilson} play two complementary roles. First, when $\gamma$ is chosen to be the smallest closed string allowed by the lattice geometry, denoted by $\gamma_{\square}$, then $W^{(\bar\alpha)}(\gamma_{\square})$ defines a local gauge-invariant operator. Such operators provide the natural pure-gauge flux terms that can be included in a local Hamiltonian. Second, for closed strings of arbitrary size, $W^{(\bar\alpha)}(\gamma)$ defines a Wilson loop of the gauged theory. These Wilson loops generate the emergent $(d-1)$-form quantum symmetry of the bond algebra $\mathfrak B_{\mathrm{iso}}$ in Eq. \eqref{iso}, denoted by $G_{\mathrm{int}}^{(d-1)}$, and measure the magnetic flux through surfaces whose boundary is $\gamma$. On lattices with periodic boundary conditions, $\gamma$ may also be chosen to be non-contractible. In this case, the interpretation in terms of flux through a bounded surface is no longer available. Nevertheless, $W^{(\bar\alpha)}(\gamma)$ remains a well-defined topological operator implementing the emergent $(d-1)$-form symmetry on the gauged theory generated by $\mathfrak B_{\mathrm{iso}}$. Finally, because the Wilson operators $W^{(\bar\alpha)}(\gamma)$ are weighted by the modulation functions $g^{(a,\bar\alpha)}$, the resulting emergent quantum symmetries inherit the modulation structure of the original internal symmetry. In this sense, they should be regarded as \textit{modulated} higher-form symmetries, rather than ordinary higher-form symmetries.


As discussed above, the gauging procedure trivializes the original symmetry operators $U^{(\alpha)}$, as shown explicitly in Eq.~\eqref{trivialization_gauge}. Nevertheless, nontrivial extended operators can still be obtained by restricting $U^{(\alpha)}$ to a finite $d$-dimensional region $\Gamma\subset\Lambda$. Using the relation between matter and gauge fields in the physical Hilbert space, Eq.~\eqref{eq:matter_gauge_constraint}, we find
\begin{eqnarray}
U^{(\alpha)}(\Gamma)
&=& \prod_{\vec r\in \Gamma}
X_{\vec r}^{f^{(\alpha)}_{\vec r}}
\nonumber\\
&=&
\prod_{a=1}^{\chi}
\prod_{\vec s_a\in \Lambda^\vee_a}
(\mathcal X_{\vec s_a}^{(a)})^{
\sum_{\vec r\in \Gamma}
\Delta^{(a)}_{\vec s,\vec r}\,
f_{\vec r}^{(\alpha)}
}.
\end{eqnarray}
For dual sites whose associated stencil lies entirely in the interior or exterior of $\Gamma$, the exponent vanishes by the symmetry constraint. More precisely,
\begin{eqnarray}
\sum_{\vec r\in \Gamma}
\Delta^{(a)}_{\vec s,\vec r}\,
f_{\vec r}^{(\alpha)}
=
0
\qquad
\text{if}
\qquad
\vec{s} \notin \partial\Gamma.
\end{eqnarray}
Therefore, only dual degrees of freedom supported near the boundary of $\Gamma$ contribute, and the truncated symmetry operator reduces to a boundary operator,
\begin{eqnarray}
\label{eqU}
U^{(\alpha)}(\Gamma)
&=&
\prod_{a=1}^{\chi}
\prod_{\vec s_a\in \partial\Gamma_a}
(\mathcal X_{\vec s_a}^{(a)})^{
\sum_{\vec r\in \Gamma}
\Delta^{(a)}_{\vec s,\vec r}\,
f_{\vec r}^{(\alpha)}
}
\nonumber\\
&\equiv&
U_{\partial}^{(\alpha)}(\partial\Gamma),
\end{eqnarray}
where $\partial\Gamma$ denotes the $(d-1)$-dimensional boundary of the region $\Gamma$ and $\partial\Gamma_a$ denotes the corresponding boundary in the $a$-th dual lattice, $\partial \Gamma_a \equiv \mu_a(\partial\Gamma)$. The subscript $\partial$ in $U_{\partial}^{(\alpha)}$ emphasizes that this operator is supported on the boundary $\partial \Gamma$ and remains nontrivial there.


\subsection{Topologically Ordered Deconfined Phase}
\label{sec: deconfined phase of gauge theory}

\begin{figure}
	\includegraphics[scale=1]{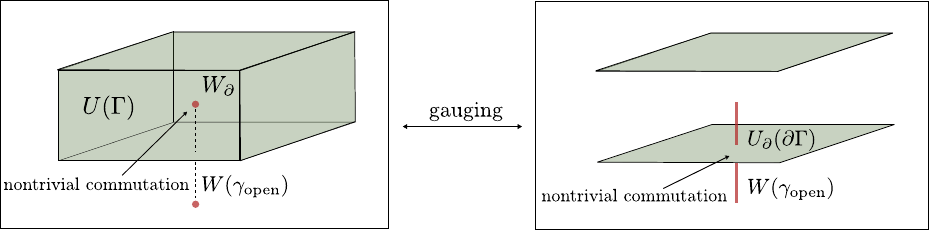}
	\caption{Geometric realization of the nontrivial algebra before gauging (left) and after gauging (right). Under gauging, operators that are nontrivial in the original theory $\mathfrak T$ become empty operators, while empty operators give rise to nontrivial operators in the gauged theory $\mathfrak T/G_{\mathrm{int}}$. }
    \label{mix_gauge}
\end{figure}

The boundary operators $U_{\partial}^{(\alpha)}(\partial\Gamma)$ become genuine $1$-form symmetry operators when we restrict to the pure-gauge subalgebra
$\mathfrak B_{\mathrm{pure\text{-}gauge}}\subset \mathfrak B_{\mathrm{iso}}$, defined by
\begin{equation}
\mathfrak B_{\mathrm{pure\text{-}gauge}}
=
\left\langle
W^{(\bar\alpha)}(\gamma_{\square}),~
\prod_{a=1}^{\chi}
\prod_{\vec s_a\in\Lambda_a^{\vee}}
\left(
\mathcal X_{\vec s_a}^{(a)}
\right)^{
\Delta^{(a)}_{\vec s,\vec r}
}
\right\rangle,
\label{nnmm}
\end{equation}
where $\gamma_{\square}$ denotes an elementary plaquette, and this subalgebra may be viewed as the result of integrating out the matter degrees of freedom from $\mathfrak B_{\mathrm{iso}}$. Since $\partial\Gamma$ is a $(d-1)$-dimensional submanifold, the operator $U_{\partial}^{(\alpha)}(\partial\Gamma)$ has the proper codimension to implement a $1$-form symmetry in any spatial dimension $d$, which we denote by $G_{\mathrm{int}}^{(1)}$. It acts naturally on the Wilson loops $W^{(\bar\alpha)}(\gamma)$, which are therefore charged under it. Indeed, they satisfy
\begin{equation}
U_{\partial}^{(\alpha)}(\partial\Gamma) W^{(\bar\alpha)}(\gamma) = \omega^{\sum_{\vec{r}_i \in (\gamma\cap \partial\Gamma)}\sum_{\vec{s} \in \Gamma} g_{\vec{r}_i}^{(a_i,\bar\alpha)}\Delta^{(a_i)}_{\vec{r}_i,\vec{s}}f_{\vec{s}}^{(\alpha)}} W^{(\bar\alpha)}(\gamma)U_{\partial}^{(\alpha)}(\partial\Gamma),
\label{mixed_gauged}
\end{equation}
which is the counterpart of \eqref{mixed} in the gauged theory. The
commutation phase is localized at the intersections between $\gamma$ and
$\partial\Gamma$. As in \eqref{crossing_global}, the choice of the local
coefficients $g_{\vec r_i}^{(a_i,\bar\alpha)}$ implies that the contribution
from these crossings satisfies
\begin{equation}
\sum_{\vec{r}_i \in (\gamma\cap \partial\Gamma)}\sum_{\vec{s} \in (\Gamma\cap B_t(\gamma \cap \partial\Gamma))} g_{\vec{r}_i}^{(a_i,\bar\alpha)}\Delta^{(a_i)}_{\vec{r}_i,\vec{s}}f_{\vec{s}}^{(\alpha)} =\mathcal{I}(\gamma, \partial\Gamma)\delta_{\alpha,\bar\alpha}
\label{crossing_gauge}.
\end{equation}
A geometric comparison between the nontrivial algebra before and after gauging is shown in Fig.~\ref{mix_gauge}.

Moreover, the two types of generators in $\mathfrak B_{\mathrm{pure\text{-}gauge}}$ commute with one another, corresponding to the gapped deconfined phase fixed-point. The nontrivial mutual action between the $1$-form symmetry generated by $U_{\partial}^{(\alpha)}(\partial\Gamma)$ and the symmetry generated by the Wilson loops $W^{(\bar\alpha)}(\gamma)$ then signals a mixed 't Hooft anomaly between the corresponding higher-form symmetries. In the pure-gauge theory in $d\geq 2$, this anomaly implies a topologically ordered phase, characterized by full quantum symmetry 
\begin{eqnarray}
G^{\vee}
=
\left(
G_{\mathrm{int}}^{(1)}
\rtimes
G_{\mathrm{int}}^{(d-1)}
\right)
\rtimes
G_{\mathrm{space}}.
\label{loppp}
\end{eqnarray}

\section{Symmetry-Enriched Topological Order}
\label{sec: SETs}

We now specialize the gauging construction developed in the previous section to two spatial dimensions and analyze the resulting symmetry-enriched topological order. In particular, we determine the action of $G_{\mathrm{space}}$ on the emergent anyonic excitations by extracting it directly from the transformation properties of the extended operators $U^{(\alpha)}(\partial\Gamma)$ and $W^{(\alpha)}(\gamma)$.

The deconfined phase of the pure gauge theory realizes an Abelian topological order described by the quantum double $\mathcal D(G_{\mathrm{int}})$. Its magnetic anyons are labeled by elements of $G_{\mathrm{int}}$, whereas its electric anyons are labeled by characters in the Pontryagin dual group $\widehat G_{\mathrm{int}}$. For the Abelian groups considered here, $G_{\mathrm{int}}=\mathbb Z_N^{\times \mathfrak q} $, we have $\widehat G_{\mathrm{int}}\simeq G_{\mathrm{int}}$. After choosing this identification and a basis, a generic anyon can therefore be labeled by a pair
\begin{equation}
    (\vec q,\vec p)\in (\mathbb Z/N\mathbb Z)^{\times \mathfrak q}\times (\mathbb Z/N\mathbb Z)^{\times \mathfrak q},
\end{equation}
where $\vec q$ and $\vec p$ are column vectors in $(\mathbb Z/N\mathbb Z)^{\times \mathfrak q}$ specifying, respectively, the electric and magnetic charges.

The braiding data are determined by the canonical pairing between $\widehat G_{\mathrm{int}}$ and $G_{\mathrm{int}}$, which, in the chosen basis, is represented by $\vec q \cdot\vec p$. Microscopically, this same pairing is encoded in the intersection algebra between the extended operators $W^{(\bar\alpha)}(\gamma)$ and $U^{(\alpha)}(\partial\Gamma)$ that create electric and magnetic excitations. Accordingly, the mutual braiding phase between the anyons $(\vec q,\vec p)$ and $(\vec q\,',\vec p\,')$ is
\begin{equation}
    M_{(\vec q,\vec p),(\vec q\,',\vec p\,')}
    =
    \omega^{\,\vec q^{\,T}\vec p\,'+\vec q\,'^{\,T}\vec p},
    \qquad
    \omega=e^{2\pi i/N}.
    \label{eq:anyon_braiding_pairing}
\end{equation}
The transformations derived below must therefore act consistently on the electric and magnetic sectors so as to preserve this pairing.


\subsection{Transformation of Magnetic Anyons}
\label{sec: transformation of magnetic anyons}

To proceed, it is convenient to rewrite the relation \eqref{transf1} as 
\begin{eqnarray}
U^g_{\mathrm{space}}\,
U^{(\alpha)}\,
U^{g\dagger}_{\mathrm{space}}
=
\prod_{\beta}
    \left[U^{(\beta)}\right]^{
    A_{\alpha,\beta}^{(g^{-1})}
    }.
    \label{eq:U_transformation_SET}
\end{eqnarray}
If a general element of $G_{\mathrm{int}}$ is represented by a column $\mathfrak q-$dimensional  vector
$\boldsymbol n \in (\mathbb Z/N\mathbb Z)^{\times \mathfrak q}$ through
\begin{equation}
    U_{\boldsymbol n}
    =
    \prod_{\alpha}
    [U^{(\alpha)}]^{n_\alpha},
\end{equation}
then Eq.~\eqref{eq:U_transformation_SET} becomes
\begin{equation}
   U^g_{\mathrm{space}}\,
U_{\boldsymbol n}\,
U^{g\dagger}_{\mathrm{space}}
    =
    U_{\varphi(g)\cdot \boldsymbol n},
\end{equation}
where $\varphi(g)$ is a $\mathfrak q\times \mathfrak q$ matrix, given by
\begin{equation}
{
    \varphi(g)
    =
    \left[A^{(g^{-1})}\right]^T
    }
    \label{eq:varphi_definition}
\end{equation}
which corresponds to a matrix representation of the homomorphism in Eq.~\eqref{eq: homomorphism varphi}.
The inverse and transpose follow from the distinction between the
transformation of the spatial modulation profiles and the active
transformation of the symmetry generators.

The boundary operators in \eqref{eqU}, which we repeat here for convenience, are
\begin{eqnarray}
U_{\partial}^{(\alpha)}(\partial\Gamma)
&=&
\prod_{a=1}^{\chi}
\prod_{\vec s_a\in \partial\Gamma_a}
(\mathcal X_{\vec s_a}^{(a)})^{
 E_{\vec s;\Gamma}^{(a,\alpha)}
},
\end{eqnarray}
where we have defined
\begin{equation}
    E_{\vec s;\Gamma}^{(a,\alpha)}
    \equiv
    \sum_{\vec r\in\Gamma}
    \Delta^{(a)}_{\vec s,\vec r}
    f_{\vec r}^{(\alpha)},
    \label{eq:E_definition}
\end{equation}
making explicit the dependence on $\Gamma$. Let us analyze the transformation of such operators under $G_{\text{space}}$. According to \eqref{gaugetransf}, we have
\begin{eqnarray}
U_{\text{space}}^g U_{\partial}^{(\alpha)}(\partial\Gamma) U_{\text{space}}^{g \dagger}
&=&\prod_{a=1}^{\chi}
\prod_{\vec s_a\in \partial\Gamma_a}
\left(\mathcal{X}_{\vec s_a^g}^{(\pi_g(a))}\right)^{\epsilon_g(a)
 E_{\vec s;\Gamma}^{(a,\alpha)}
}. \label{9990}
\end{eqnarray}
Using \eqref{rep} and \eqref{eq:Delta_covariance}, we can express
\begin{eqnarray}
E_{\vec s;\Gamma}^{(a,\alpha)}&=& \epsilon_g(a)\sum_{\beta}\sum_{\vec r\in\Gamma}  \Delta^{(\pi_g(a))}_{\vec{s}^g,\vec{r}^g} A^{(g^{-1})}_{\alpha,\beta}f_{\vec{r}}^{(\beta)}\nonumber\\
&=&\epsilon_g(a)\sum_{\beta}E_{\vec{s}^g;\Gamma^g}^{(\pi_g(a), \beta)}A^{(g^{-1})}_{\alpha,\beta}.
\end{eqnarray}
Plugging this form in \eqref{9990}, we obtain 
\begin{equation}
U_{\text{space}}^g U_{\partial}^{(\alpha)}(\partial\Gamma) U_{\text{space}}^{g \dagger}=  \prod_{\beta}
    \left[
    U_{\partial}^{(\beta)}(\partial\Gamma^g)
    \right]^{
    A_{\alpha\beta}^{(g^{-1})}
    }.
    \label{eq:V_transformation}
\end{equation}

In two spatial dimensions, open counterparts of the
$U_{\partial}$-operators create magnetic anyons at their endpoints. Therefore
Eq.~\eqref{eq:V_transformation} implies
\begin{equation}
  g:  m_\alpha 
    \rightarrow
    \prod_{\beta}
    m_\beta^{
    A_{\alpha\beta}^{(g^{-1})}
    }.
    \label{eq: flux anyons transformation}
\end{equation}
Equivalently, if a general magnetic-flux label is represented by a column vector $\vec p$, then
\begin{equation}
    \vec p
    \rightarrow
    \varphi(g)\vec p
    =
    (A^{(g)})^{-T}\vec p.
    \label{eq:magnetic_column_transformation}
\end{equation}
Thus, the action of $G_{\mathrm{space}}$ on the magnetic anyons is directly inherited from the automorphism $\varphi(g)$ induced on $G_{\mathrm{int}}$.


\subsection{Transformation of Electric Anyons} \label{sec: transformation of electric anyons}

Now we turn to the transformation of electric anyons, which are described in terms of the Wilson operators \eqref{Wilson}. In this way, we need to understand the transformation property of the functions $g_{\vec{r}}^{(a,\bar\alpha)}$ under $G_{\text{space}}$. We consider a generic transformation 
\begin{equation}
g:\, g_{\vec{r}}^{(a,\bar\alpha)} \rightarrow g_{\vec{r}^g}^{(a,\bar\alpha)}=\sum_{\bar\beta}C^{(g)}_{\bar\alpha,\bar\beta}\sum_b M^{(g)}_{a,b} g_{\vec{r}}^{(b,\bar\beta)},
\end{equation}
where the index $a$ is assumed to transform according to the monomial
representation in Eq.~\eqref{monomial}, while the matrices $C^{(g)}$ remain
to be determined. Thus, the transformation above can be written as
\begin{equation}
g^{(\pi_g(a),\bar\alpha)}_{\vec{r}^g}=\epsilon_g(a)\sum_{\bar\beta} C^{(g)}_{\bar\alpha,\bar\beta} g^{(a,\bar\beta)}_{\vec{r}}.    
\end{equation}

To determine $C^{(g)}$, we consider the bilinear pairing that characterizes the intersection algebra of the extended operators in Eq.~\eqref{crossing_gauge} and examine its transformation under $g$, namely,
\begin{equation}
\sum_{\vec{r}_i \in (\gamma\cap \partial\Gamma)}\sum_{\vec{s} \in (\Gamma\cap B_t(\gamma \cap \partial\Gamma))}  g_{\vec{r}_i^g}^{(\pi_g(a_i),\bar\alpha)}\Delta^{(\pi_g(a_i))}_{\vec{r}_i^g,\vec{s}^g}f_{\vec{s}^g}^{(\alpha)}= \mathcal{I}(\gamma^g,\partial\Gamma^g) \delta_{\alpha,\bar\alpha}.
\end{equation}
Using the transformation properties of $g_{\vec r}^{(a,\bar\alpha)}$, $f_{\vec r}^{(\alpha)}$, and $\Delta^{(a)}_{\vec r,\vec s}$, together with $\epsilon_g(a)^2=1$, we obtain
\begin{eqnarray}
\mathcal{I}(\gamma^g,\partial\Gamma^g) \delta_{\alpha,\bar\alpha} = \left(C^{(g)}(A^{(g)})^T \right)_{\alpha,\bar\alpha}  \mathcal{I}(\gamma,\partial\Gamma).
\end{eqnarray}
Since the spatial transformation maps the pair $(\gamma,\partial\Gamma)$ onto $(\gamma^g,\partial\Gamma^g)$ while preserving their intersection structure, the corresponding intersection index is invariant,
\begin{equation}
\mathcal{I}(\gamma,\partial\Gamma)=\mathcal{I}(\gamma^g,\partial\Gamma^g).
\end{equation}
It then follows that
\begin{equation}
C^{(g)} = (A^{(g)})^{-T}.
\end{equation}
Therefore, the functions $g_{\vec r}^{(a,\bar\alpha)}$ transform according to the contragredient representation of $A^{(g)}$,
\begin{equation}
g:\, g_{\vec{r}}^{(a,\bar\alpha)} \rightarrow g_{\vec{r}^g}^{(\pi_g(a),\bar\alpha)}=\epsilon_g(a)\sum_{\bar\beta}(A^{(g)})^{-T}_{\bar\alpha,\bar\beta}\, g_{\vec{r}}^{(a,\bar\beta)}.
\label{gtransf}
\end{equation}


With the transformation law \eqref{gtransf} in hand, we can now determine the action of $G_{\text{space}}$ on the Wilson operators \eqref{Wilson}. Under a spatial transformation $g\in G_{\text{space}}$, the Wilson operator transforms as
\begin{eqnarray}
U^g_{\text{space}}W^{(\bar\alpha)}(\gamma) U^{g \dagger}_{\text{space}}&=&
\prod_{i=1}^n
\left(
\mathcal Z_{\vec r_{a_i}^g}^{(\pi_g(a_i))}
\right)^{\epsilon_g(a_i) g^{(a_i,\bar\alpha)}_{\vec{r}_i}}.
\end{eqnarray}
To express the right-hand side in terms of Wilson operators defined on the transformed string $\gamma^g$, we use \eqref{gtransf} in the inverse form
\begin{equation}
g_{\vec{r}}^{(a,\bar\alpha)}=\epsilon_g(a)\sum_{\bar\beta}(A^{(g^{-1})})^{-T}_{\bar\alpha,\bar\beta}\, g_{\vec{r}^g}^{(\pi_g(a),\bar\beta)},
\end{equation}
which gives
\begin{eqnarray}
U^g_{\text{space}}W^{(\bar\alpha)}(\gamma) U^{g \dagger}_{\text{space}}&=&\prod_{\bar\beta}\left[
\prod_{i=1}^n
\left(
\mathcal Z_{\vec r_{a_i}^g}^{(\pi_g(a_i))}
\right)^{g^{(\pi_g(a_i),\bar\beta)}_{\vec{r}_i^g}}\right]^{(A^{(g^{-1})})_{\bar\alpha,\bar\beta}^{-T}}\nonumber\\
&=& \prod_{\bar\beta} \left[ W^{(\bar\beta)}(\gamma^g) \right]^{(A^{(g^{-1})})_{\bar\alpha,\bar\beta}^{-T}}.
\end{eqnarray}
Thus, for an open string, the electric anyons located at its endpoints transform according to
\begin{eqnarray}
  g:  e_{\bar\alpha} 
    &\rightarrow&
    \prod_{\beta}
    e_{\bar\beta}^{
    [A^{(g^{-1})}]^{-T}_{\bar\alpha\bar\beta}
    }\nonumber\\
    &=& \prod_{\bar\beta}
    e_{\bar\beta}^{
    [A^{(g)}]^{T}_{\bar\alpha\bar\beta}
    },
    \label{eq: charge anyons transformation}
\end{eqnarray}
where in the last step we used $A^{(g^{-1})}=(A^{(g)})^{-1}$. Therefore, the electric-anyon labels transform according to the transpose representation $(A^{(g)})^{T}$.



\subsection{Generic Anyon Transformation}\label{sec: generic anyon transformation}

Let us now consider a generic anyon parametrized as
\begin{equation}
\mathsf{a}(\vec{q},\vec{p}) \equiv \left(\prod_{\bar\alpha=1}^{\mathfrak{q}} e_{\bar\alpha}^{q_{\bar\alpha}}\right)\left( \prod_{\alpha=1}^q m_{\alpha}^{p_{\alpha}}\right).  
\end{equation}
Combining the transformations in \eqref{eq: flux anyons transformation} and \eqref{eq: charge anyons transformation}, we find
\begin{equation}
g: \mathsf{a}(\vec{q},\vec{p}) ~\rightarrow~ \mathsf{a}(\vec{q}',\vec{p}'), 
\end{equation}
with
\begin{equation}
\vec{q}'= A^{(g)}\vec{q} ~~~\text{and}~~~ \vec{p}'=(A^{(g)})^{-T} \vec{p}.
\end{equation}
Thus, the electric and magnetic charge vectors transform in mutually contragredient representations. The complete action of $G_{\text{space}}$ on the anyon labels can therefore be written as
\begin{equation}
g:\begin{pmatrix}
\vec{q}\\
\vec{p}
\end{pmatrix}
~\rightarrow~
\begin{pmatrix}
\vec{q}'\\
\vec{p}'
\end{pmatrix}=
\begin{pmatrix}
A^{(g)} & 0\\
0 & (A^{(g)})^{-T}
\end{pmatrix}
\begin{pmatrix}
\vec{q}\\
\vec{p}
\end{pmatrix}.
\end{equation}

The mutually contragredient action on the electric and magnetic sectors has a direct topological interpretation. It preserves each electric-magnetic pairing entering Eq.~\eqref{eq:anyon_braiding_pairing}, and hence leaves the full mutual braiding phases and topological spins invariant. Thus, for the orientation-preserving spatial symmetries considered here, the resulting anyon permutation defines a braided autoequivalence of the quantum-double topological order. The appearance of $A^{(g)}$ and $(A^{(g)})^{-T}$ is therefore not only a consequence of the microscopic gauging construction, but precisely the structure required for compatibility with the braiding data.

The discussion above is formulated at the level of the local operator algebra and is therefore insensitive to global properties of the lattice. On a compact lattice, however, periodic boundary conditions can impose additional global relations among the Wilson operators, potentially modifying their orders or reducing the number of independent electric-magnetic pairs. Throughout this work, we avoid such finite-size effects by choosing system sizes compatible with all modulation profiles under consideration.

Several extensions of the present construction are possible. Introducing multiple onsite species, additional sublattices, or more general transformations among the gauge-field flavors enlarges the class of spatial symmetry representations that can be realized by the modulation profiles. Moreover, stacking these phases with $G_{\mathrm{space}}$-protected SPT phases \cite{Han:2023fas} produces SETs with the same intrinsic topological order, anyon permutation, and symmetry-fractionalization data, while modifying the associativity data of the symmetry defects. More general symmetry actions that mix electric and magnetic anyons would require a microscopic realization of electric-magnetic duality, possibly involving an enlarged unit cell or additional microscopic degrees of freedom. Finally, incorporating orientation-reversing crystalline symmetries requires extending the present framework to symmetry actions described by anti-braided autoequivalences \cite{Barkeshli:2016mew,Barkeshli:2019vtb}.



\section{One-Dimensional Chain}\label{1D}

Although modulated symmetries admit infinitely many distinct modulation patterns, we can nevertheless identify broad, systematic classes of them by focusing on sufficiently low-dimensional representations $\mathfrak{q}$. In this and in the following section, we explore in some generality the bond algebras that arise from such low-dimensional induced representations $A: G_{\mathrm{space}} \to \mathrm{GL}(\mathfrak{q}, \mathbb{Z}_N)$.
We shall start with the one-dimensional lattice, where the relevant spatial symmetry group reduces to lattice translations,
\begin{equation}
G_{\text{space}}=T\cong \mathbb Z.
\end{equation}


\subsection{$\mathfrak{q}=1$, $\mathbb{Z}_N$}

The lowest-dimensional induced representation is obtained with $\mathfrak{q}=1$, so that $G_{\text{int}}$ contains a single $\mathbb{Z}_N$ factor. The total symmetry group is then $G=  \mathbb{Z}_N \rtimes T$.
In this case, the symmetry is specified by a single function $f_i$, and Eq.~\eqref{rep} reduces to
\begin{equation}
f_{i^T}=f_{i+1}=A f_i,
\end{equation}
where $A\in \mathbb Z/N\mathbb Z$  and $T$ denotes the translation generator. This gives a simple recurrence relation, whose solution is
\begin{equation}
f_i = A^i f_0.
\end{equation}
This corresponds to the class of exponentially modulated symmetries~\cite{Watanabe:2022pgk,PhysRevB.107.195139, Delfino:2023rpw,Lian:2022nqj,PhysRevB.110.165109,ryb5-8ntp,Lian:2026rwk}. Choosing $f_0=1$, the associated symmetry operator takes the form
\begin{equation}
U=\prod_i (X_i)^{A^i}.
\label{oned}
\end{equation}
The uniform case is recovered by setting $A=1$.


Now we can proceed to determine the bond-algebra by following the construction described in \eqref{3430},
\begin{equation}
\sum_{j\in B_t(i)} F_{1,j}K_{i-j} =0 ~~~
    \mathrm{mod}\, N,
\end{equation}
with $F_{1,i}=A^{i}$.
The minimal stencil is $t=2$, which implies $j=i,i+1$. In this case, this equation reduces to 
\begin{equation}
F_{1,i}K_{i-i} + F_{1,i+1}K_{i-(i+1)} =0 ~~~
    \mathrm{mod}\, N.
\end{equation}
\begin{equation}
A^i(K_0 + A K_{-1})=0 ~~~
    \mathrm{mod}\, N,
\end{equation}
whose solution is $K_0 = - A K_{-1} ~~\mathrm{mod}\, N$. Setting $K_{-1}=-1$, we obtain $K_0=A~~ \mathrm{mod}\, N$. In terms of $\Delta_{i,j}$, we have $\Delta_{i,j}= A\delta_{i,j}-\delta_{i+1,j}$. The corresponding symmetric bond-algebra is
\begin{equation}
 \mathfrak{B}_{\mathrm{matter}} =\left\langle Z_i^AZ_{i+1}^{\dagger},~X_i \right\rangle.
\end{equation}

In the uniform case $A=1$, this bond-algebra reduces to the usual $\mathbb{Z}_N$ Ising bond-algebra.

\subsection{$\mathfrak{q}=2$, $\mathbb{Z}_N\times \mathbb{Z}_N$}

Next we consider the case $\mathfrak{q}=2$, for which the relevant symmetry group is $G=  (\mathbb{Z}_N\times \mathbb{Z}_N) \rtimes T$.
In this case, Eq.~\eqref{rep} becomes
\begin{equation}
\begin{pmatrix}
    f_{i+1}^{(1)}  \\
    f_{i+1}^{(2)} 
\end{pmatrix}= 
\begin{pmatrix}
   A_{1,1} & A_{1,2} \\
   A_{2,1} & A_{2,2}
\end{pmatrix}
\begin{pmatrix}
    f_{i}^{(1)}  \\
    f_{i}^{(2)} 
\end{pmatrix}.
\end{equation}

We now construct the matrix $F_{\alpha,j}=f_j^{(\alpha)}$ using the minimal stencil $t=3$, with $j=i, i+1, i+2$. This choice ensures the existence of a nontrivial kernel for any initial condition of the recurrence relation. The kernel 
\begin{equation}
\sum_{\beta}(K_0 \delta_{\alpha,\beta}+K_{-1}A_{\alpha,\beta}+K_{-2}A^2_{\alpha,\beta})f^{(\beta)}_i=0 \mod N,
\end{equation}
follows from the Cayley-Hamilton characteristic equation
\begin{equation}
A^2- (\text{Tr}A) A + \det A =0.
\end{equation}
Thus, the kernel is generated by
\begin{equation}
\ker{F}= (-\det A,\text{Tr}A,-1) 
\label{ker001}
\end{equation}
and the corresponding bond algebra reads
\begin{equation}
 \mathfrak{B}_{\mathrm{matter}} =\left\langle (Z_i)^{-\det A}(Z_{i+1})^{\text{Tr}A} Z_{i+2}^{\dagger},~X_i \right\rangle.
\end{equation}

It is useful to distinguish a few particular choices of the matrix $A$ that lead to familiar classes of modulated symmetries. First, when $A_{1,1}=A_{2,2}=0$, the recurrence relation takes the off-diagonal form
\begin{equation}
f_{i+1}^{(1)}=A_{1,2}f_i^{(2)},
\qquad
f_{i+1}^{(2)}=A_{2,1}f_i^{(1)} .
\end{equation}
In this case, translation by one lattice spacing exchanges the two components. After two translations, each component evolves independently according to
\begin{equation}
f_{i+2}^{(1)}=A_{1,2}A_{2,1}f_i^{(1)},
\qquad
f_{i+2}^{(2)}=A_{1,2}A_{2,1}f_i^{(2)} .
\end{equation}
Thus the symmetry naturally decomposes into even and odd sublattices, with an effective exponential modulation controlled by $A_{1,2}A_{2,1}$ under translations by two sites.

Second, when $A_{1,2}=A_{2,1}=0$, the two components decouple:
\begin{equation}
f_{i+1}^{(1)}=A_{1,1}f_i^{(1)},
\qquad
f_{i+1}^{(2)}=A_{2,2}f_i^{(2)} .
\end{equation}
This gives two independent exponential modulated symmetries, one for each $\mathbb Z_N$ factor in $G_{\mathrm{int}}$. The previous $\mathfrak q=1$ example is recovered by focusing on either component separately.

Finally, the case $A_{1,1}=0$ gives
\begin{equation}
f_{i+1}^{(1)}=A_{1,2}f_i^{(2)},
\qquad
f_{i+1}^{(2)}=A_{2,1}f_i^{(1)}+A_{2,2}f_i^{(2)} .
\end{equation}
In this case, the first component at the next site is determined by the second component at the previous site, while the second component mixes both fields. Equivalently, eliminating $f_i^{(2)}$ gives a second-order recurrence for $f_i^{(1)}$,
\begin{equation}
f_{i+2}^{(1)}=A_{2,2}f_{i+1}^{(1)}+A_{1,2}A_{2,1}f_i^{(1)},
\end{equation}
assuming $A_{1,2}$ is invertible in $\mathbb Z_N$. This form makes explicit that the modulation is not simply exponential in one step, but is instead governed by a two-step recurrence.


\section{Two-Dimensional Square Lattice}\label{2D}

In two spatial dimensions, rotations can also play a nontrivial role in the structure of the modulation. Although our construction holds, in principle, for arbitrary lattices, here we focus on the square lattice, whose spatial symmetry group is $G_{\text{space}}=T^2 \rtimes C_4$, where the semidirect product structure reflects the nontrivial action of rotations on translations. Denoting by $T_x$ and $T_y$ the translation generators and by $R$ the generator of $C_4$, this action is encoded in the relations
\begin{equation}
R\, T_x \,R^{-1} = T_y ~~~\text{and}~~~ R\, T_y\, R^{-1} = T_x^{-1}.
\label{c41}
\end{equation}
In addition, we have
\begin{equation}
R^4=1.
\label{c42}
\end{equation}

If the model is anisotropic between the $x$ and $y$ directions, the $C_4$ rotation symmetry is generally broken. In this case $G_{\mathrm{space}} = T^2$, and  the modulated structure is essentially governed by translations alone.


\subsection{$\mathfrak{q}=1$, $\mathbb{Z}_N$}

\subsubsection{Anisotropic Case}

Let us begin with the anisotropic case, for which the relation \eqref{rep} gives
\begin{eqnarray}
f_{\vec{r}+\hat{x}}&=&A^{(T_x)} f_{\vec{r}};\nonumber\\
f_{\vec{r}+\hat{y}}&=&A^{(T_y)}f_{\vec{r}},
\label{rani}
\end{eqnarray}
where we parameterize the lattice position as $\vec{r}= i_x \hat{x}+ i_y \hat{y} = (i_x,i_y)$.

We proceed by constructing the matrix $F_{1,{\vec{s}}}=f^{(1)}_{\vec{s}}\equiv f_{\vec{s}}$ using the minimal stencil $t=3$, involving the sites $\vec{s}=\vec{r},\,\vec{r}+\hat{x},\, \vec{r}+\hat{y}$. The corresponding kernel is generated by
\begin{equation}
\ker{F}= \left\langle (A^{(T_x)},-1,0), (A^{(T_y)},0,-1)\right\rangle.
\end{equation}
This leads to the bond-algebra
\begin{equation}
\mathfrak{B}_{\mathrm{matter}} =\left\langle Z_{\vec{r}}^{A^{(T_x)}}Z_{\vec{r}+\hat{x}}^{\dagger}, ~Z_{\vec{r}}^{A^{(T_y)}}Z_{\vec{r}+\hat{y}}^{\dagger}, ~X_{\vec{r}} \right\rangle,
\label{baq1}
\end{equation}
which corresponds to independent exponential modulations along the two lattice directions.


\subsubsection{Isotropic Case: $C_4$-Invariant}

We now turn to the isotropic case with the $C_4$ symmetry. In addition to the translation relations in Eq.~\eqref{rani}, the function $f_{\vec r}$ must also transform under the $\pi/2$-rotation, 
\begin{eqnarray}
f_{\vec{r}^R}= A^{(R)} f_{\vec{r}},
\label{111}
\end{eqnarray}
with $\vec{r}^R=(-i_y,i_x)$. Compatibility between translations and rotations, together with the square-lattice relation $R^4=1$, then implies
\begin{equation}
A^{(T_x)}=A^{(T_y)}~~~\text{and}~~~ A^{(T_y)}=(A^{(T_x)})^{-1}~ \Rightarrow~ (A^{(T)})^2=1\mod N
\end{equation}
and
\begin{equation}
(A^{(R)})^4=1 \mod N.
\end{equation}
The recurrence relations further constrain the parameters, 
\begin{equation}
A^{(T_x)}=A^{(T_y)}=\pm 1 \mod N~~~\text{and}~~~ A^{(R)}=1 \mod N.
\end{equation}
Thus, we obtain $f_{\vec{r}}=(\pm 1)^{i_x+i_y}$. The corresponding bond-algebras follow from Eq.~\eqref{baq1} by setting $A^{(T_x)}=A^{(T_y)}=\pm 1$.


\subsection{$\mathfrak{q}=2$, $\mathbb{Z}_N\times \mathbb{Z}_N$}

\subsubsection{Anisotropic Case}

We start by imposing only translation covariance,
\begin{eqnarray}
f^{(\alpha)}_{\vec{r}+\hat{x}}&=&\sum_{\beta}A_{\alpha,\beta}^{(T_x)} f^{(\beta)}_{\vec{r}};\nonumber\\
f^{(\alpha)}_{\vec{r}+\hat{y}}&=&\sum_{\beta}A_{\alpha,\beta}^{(T_y)} f^{(\beta)}_{\vec{r}},
\label{22312}
\end{eqnarray}
where $A^{(T_x)}$ and $A^{(T_y)}$ are $2\times 2$ matrices, since $\alpha,\beta=1,2$.

Instead of proceeding in full generality, we parametrize $A^{(T_y)}$ as
\begin{equation}
A^{(T_y)} = b_01  + b_1 A^{(T_x)},~~~ b_1\neq 0,
\label{5656}
\end{equation}
which is sufficient to capture the relevant examples discussed below.

We then construct the matrix $F_{\alpha,\vec{s}}$. In this case, the minimal stencil involves $t=4$ sites, and different choices of its shape are possible. For instance, if we require the sites in the stencil to be separated by the shortest lattice distances, we may choose $\vec{s}=\vec{r},\,\vec{r}+\hat{x},\,\vec{r}+\hat{y},\,\vec{r}+\hat{x}+\hat{y}$. The kernel is then generated by
\begin{equation}
\ker{F} =  \left\langle (b_0, b_1, -1,0), (-b_1 \delta,(b_0+b_1\tau),0,-1) \right\rangle,
\end{equation}
where we have defined
\begin{equation}
\delta\equiv \det A^{(T_x)} ~~~\text{and}~~~ \tau\equiv  \text{Tr}A^{(T_x)}.
\end{equation}
The corresponding bond algebra is
\begin{equation}
\mathfrak{B}_{\mathrm{matter}} =\left\langle Z_{\vec{r}}^{b_0}Z_{\vec{r}+\hat{x}}^{b_1}Z_{\vec{r}+\hat{y}}^{\dagger},  ~Z_{\vec{r}}^{-b_1\delta}Z_{\vec{r}+\hat{x}}^{b_0+b_1\tau} Z_{\vec{r}+\hat{x}+\hat{y}}^{\dagger}, ~X_{\vec{r}} \right\rangle.
\end{equation}

Alternatively, by allowing second-neighbor sites in the stencil, we may consider $\vec{s}=\vec{r}-\hat{x},\,\vec{r},\,\vec{r}+\hat{x},\,\vec{r}+\hat{y}$, which is valid in the anisotropic case. The kernel is generated by
\begin{equation}
\ker{F} = \left\langle (-\delta,\tau,-1,0), (0,b_0,b_1,-1) \right\rangle.
\end{equation}
The corresponding bond algebra is
\begin{equation}
\mathfrak{B}_{\mathrm{matter}} =\left\langle Z_{\vec{r}-\hat{x}}^{-\delta}Z_{\vec{r}}^{\tau}Z_{\vec{r}+\hat{x}}^{\dagger},  ~Z_{\vec{r}}^{b_0}Z_{\vec{r}+\hat{x}}^{b_1} Z_{\vec{r}+\hat{y}}^{\dagger}, ~X_{\vec{r}} \right\rangle.
\end{equation}



\subsubsection{Isotropic Case: $C_4$-Invariant}

We now discuss the case with $C_4$ rotational invariance. In addition to the translational recurrence relations, we impose
\begin{eqnarray}
\label{517}
f^{(\alpha)}_{\vec{r}^R}=\sum_{\beta} A^{(R)}_{\alpha,\beta} f^{(\beta)}_{\vec{r}},
\end{eqnarray}
with 
\begin{equation}
(A^{(R)})^4=1 \mod N.
\label{6.18}
\end{equation}
The matrices $A^{(T_x)}$, $A^{(T_y)}$, and $A^{(R)}$ must satisfy the compatibility conditions
\begin{equation}
A^{(R)}A^{(T_x)} {A^{(R)}}^{-1} = A^{(T_y)}~~~\text{and}~~~ A^{(R)}A^{(T_y)} {A^{(R)}}^{-1} = {A^{(T_x)}}^{-1},
\label{cond}
\end{equation}
which imply
\begin{equation}
\text{Tr}A^{(T_x)}=\text{Tr}A^{(T_y)}\equiv \tau~~~ \text{and}~~~ \det A^{(T_x)} =\det A^{(T_y)} \equiv \delta, ~\text{with}~\delta^2=1.
\end{equation}
Using the Cayley-Hamilton equation together with the above conditions, we obtain 
\begin{equation}
(1-\delta)\tau =0.    
\end{equation}
Thus, there are two possibilities:
\begin{equation}
\delta=1, \tau ~\text{arbitrary}~~~\text{or}~~~ \delta=-1, \tau=0.
\end{equation}

We again use the parametrization of $A^{(T_y)}$ introduced in Eq.~\eqref{5656}, which we repeat here for convenience,
\begin{equation}
A^{(T_y)} = b_01  + b_1 A^{(T_x)}.
\label{5656b}
\end{equation}
The above relations lead to the constraints
\begin{equation}
 b_0^2+b_0 b_1 \tau +(b_1^2-1)\delta =0 ~~~\text{and}~~~ 2b_0+(b_1-1)\tau =0. 
\end{equation}
If $\tau^2-4\delta \neq 0$, two particular solutions are $b_0=0$ and $b_1=\pm1 \mod N$. In this case, there are two possibilities:
\begin{equation}
A^{(T_y)} =\pm  A^{(T_x)}.
\label{6.25}
\end{equation}
Substituting these solutions into Eq.~\eqref{cond} further gives
\begin{equation}
(A^{(T_x)})^2 =(A^{(T_y)})^2= 1\mod N. 
\end{equation}

We now consider the stencil with $t=4$, involving the sites $\vec{s}=\vec{r},\,\vec{r}+\hat{x},\,\vec{r}+\hat{y},\,\vec{r}+\hat{x}+\hat{y}$.
With this stencil, the generators of the kernel are
\begin{equation}
\ker{F} = \left\langle (1,0,0,-1),\, (0,1,-1,0)\right\rangle.
\end{equation}
The corresponding bond algebra reads
\begin{equation}
\label{528}
\mathfrak{B}_{\mathrm{matter}} =\left\langle Z_{\vec{r}}Z_{\vec{r}+\hat{x}+\hat{y}}^{\dagger}, ~Z_{\vec{r}+\hat{x}} Z_{\vec{r}+\hat{y}}^{\dagger}, ~X_{\vec{r}}\right\rangle.
\end{equation}



\section{Explicit Realization of an Isotropic SET}
\label{VI}

We now use the isotropic $d=2$ case discussed above, with $\mathfrak{q}=2$ and $G_{\mathrm{space}}=T^2\rtimes C_4$, as a prototype to illustrate the formalism developed in this work. Gauging the modulated symmetries yields an explicit construction of an isotropic SET phase described by the quantum double $\mathcal{D}(\mathbb{Z}_N\times\mathbb{Z}_N)$. Its nontrivial crystalline enrichment is determined by the $\mathbb{Z}_2$ image of the translation group, generated by the action of a unit translation along either the $x$ or $y$ direction, which exchanges the two $\mathbb{Z}_N$ factors.

\subsection{Global Modulated Symmetries}

A representative choice of modulation functions is
\begin{equation}
f^{(1)}_{i_x,i_y} = \frac12(1+(-1)^{i_x+i_y+1})
\quad \text{and} \quad
f^{(2)}_{i_x,i_y}= \frac12(1+(-1)^{i_x+i_y}).
\end{equation}
Here, the parity of a site $\vec{r}=i_x \hat{x}+ i_y \hat{y}$ is determined by the parity of $i_x+i_y$. Accordingly, $f^{(1)}_{i_x,i_y}$ is nonzero only on the odd sublattice, whereas $f^{(2)}_{i_x,i_y}$ is nonzero only on the even sublattice. The corresponding symmetry operators can also be expressed with support restricted to the respective sublattices
\begin{equation}
U^{(1)}= \prod_{\vec{r} \in \text{odd}} X_{\vec{r}}~~~\text{and}~~~ U^{(2)}= \prod_{\vec{r} \in \text{even}} X_{\vec{r}},
\end{equation}
with $(U^{(1)})^N=1$ and $(U^{(2)})^N=1$. 

We next determine how these modulation functions transform under the generators of the space group. The corresponding representation matrices are
\begin{equation}
A^{(T_x)} = A^{(T_y)}=
\begin{pmatrix}
0 & 1 \\
1 & 0
\end{pmatrix}~~~\text{and}~~~ A^{(R)}=1,
\label{bnbn}
\end{equation}
which satisfy the consistency conditions \eqref{6.18}, \eqref{cond}, and \eqref{6.25}. Thus, translations along either lattice direction exchange the two internal symmetry generators, while rotations leave each of them invariant.

For concreteness, we consider a $G_{\mathrm{space}}$-symmetric  Hamiltonian belonging to the bond algebra  in Eq.~(\ref{528}):
\begin{equation}
\label{A64}
H = -\sum_{\vec{r}} \left(
Z_{\vec{r}}Z^\dagger_{\vec{r}+\hat{x}+\hat{y}}
+ Z_{\vec{r}+\hat{x}}Z^\dagger_{\vec{r}+\hat{y}}
\right)
- \lambda \sum_{\vec{r}}X_{\vec{r}}
+ \mathrm{H. c.}.
\end{equation}
The local operators $Z_{\vec{r}}$ on the odd and even sublattices are charged under $U^{(1)}$ and $U^{(2)}$, respectively,
\begin{equation}
U^{(1)} Z_{\vec{r}\,\in\, \text{odd}}=\omega Z_{\vec{r}\,\in\, \text{odd}} \, U^{(1)} ~~~\text{and}~~~ U^{(2)} Z_{\vec{r}\,\in\, \text{even}}=\omega Z_{\vec{r}\,\in\, \text{even}}\, U^{(2)}.
\label{ssb4}
\end{equation}
These relations are inherited from the algebra of open operators derived in Eq.~\eqref{mixed}. In the regime $\lambda\ll 1$, the system enters a spontaneous symmetry-breaking phase in which the local operators $Z_{\vec{r}}$ generate emergent two-form symmetries. Together with the original zero-form symmetries, they realize the mixed 't Hooft anomalies given in Eq.~\eqref{ssb4}. The complete symmetry structure is therefore
\begin{equation}
G=\left((\mathbb{Z}_N^{(0)}\times \mathbb{Z}_N^{(0)}) \rtimes (\mathbb{Z}_N^{(2)}\times \mathbb{Z}_N^{(2)})\right)\rtimes
G_{\mathrm{space}}.
\end{equation}


\subsection{Gauging the Modulated Symmetries}

\begin{figure}
    \centering
    \includegraphics[width=0.7\linewidth]{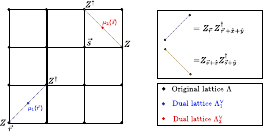}
    \caption{Operators of the bond algebra in Eq. (\ref{528}). The dual lattice $\Lambda_1^\vee$ contains the blue dots at the center of plaquettes and $\Lambda_2^\vee$ contains red dots.}
    \label{fig9}
\end{figure}

To gauge  $G_{\mathrm{int}} = \mathbb Z_N^{(0)}\times \mathbb Z_N^{(0)}$, we first identify the dual lattices associated with the operators generating the bond algebra. The two types of bond operators give rise to dual sublattices located at the centers of the plaquettes. The first is defined by
\begin{eqnarray}
    \mu_1: \Lambda &\longrightarrow& \Lambda_1^\vee \\
    \nonumber
    \vec{r} &\longmapsto& \vec{r}+\frac{1}{2}(\hat{x}+\hat{y}),
\end{eqnarray}
with fields denoted by
\begin{equation}
\mathcal{X}^{(1)}_{\vec{r}+\frac{1}{2}(\hat{x}+\hat{y})} ~~~\text{and}~~~ \mathcal{Z}^{(1)}_{\vec{r}+\frac{1}{2}(\hat{x}+\hat{y})},
\end{equation}
whereas the second sublattice is defined by
\begin{eqnarray}
    \mu_2: \Lambda &\longrightarrow& \Lambda_2^\vee \\
    \nonumber
    \vec{r} &\longmapsto& \vec{r}+\frac{1}{2}(\hat{x}+\hat{y}),
\end{eqnarray}
and hosts gauge degrees of freedom denoted by
\begin{equation}
\mathcal{X}^{(2)}_{\vec{r}+\frac{1}{2}(\hat{x}+\hat{y})} ~~~\text{and}~~~ \mathcal{Z}^{(2)}_{\vec{r}+\frac{1}{2}(\hat{x}+\hat{y})}.
\end{equation}
Thus, although the dual sublattices $\Lambda_1^\vee$ and $\Lambda_2^\vee$ coincide geometrically, as illustrated in Fig.~\ref{fig9}, the gauge degrees of freedom supported on them carry distinct flavors associated with the two types of bond operators.
Accordingly, the enlarged Hilbert space is
\begin{equation}
\mathcal{H}_{\text{matter}}\otimes \mathcal{H}_{\text{gauge}}
=
\bigotimes_{\vec{r} \in \Lambda}\mathbb{C}_{\vec{r}}^N
\otimes
\bigotimes_{a=1,2}
\bigotimes_{\vec{r}_a \in \Lambda_a^\vee}
\mathbb{C}_{\vec{r}_a}^N.
\end{equation}

From the bond algebra in Eq.~\eqref{528}, we identify, up to overall signs,
\begin{equation}
\Delta^{(1)}_{\vec r,\vec s}
=\delta_{\vec r,\vec s}
-\delta_{\vec r+\hat x+\hat y,\vec s},
\qquad
\Delta^{(2)}_{\vec r,\vec s}
=\delta_{\vec r+\hat y,\vec s}
-\delta_{\vec r+\hat x,\vec s}.
\label{zaq}
\end{equation}
Using Eq.~\eqref{eq:matter_gauge_constraint}, these coefficients give the Gauss-law operator at each site $\vec r$:
\begin{equation}
G_{\vec r}
=
X_{\vec r}^{\dagger}
\left(\mathcal X^{(1)}_{\vec r-\frac12(\hat x+\hat y)}\right)^{\dagger}
\mathcal X^{(1)}_{\vec r+\frac12(\hat x+\hat y)}
\left(\mathcal X^{(2)}_{\vec r-\frac12(\hat x-\hat y)}\right)^{\dagger}
\mathcal X^{(2)}_{\vec r+\frac12(\hat x-\hat y)}.
\end{equation}
The physical Hilbert space consists of states satisfying
\begin{equation}
\mathcal H_{\mathrm{phys}}
=
\bigl\{\,|\psi\rangle\in\mathcal H_{\mathrm{tot}}
\mid G_{\vec r}|\psi\rangle=|\psi\rangle
\text{ for every }\vec r\,\bigr\}.
\end{equation}


The gauge-invariant bond algebra is generated by the minimally coupled bond operators obtained from Eq.~(\ref{528}), together with the operators $X_{\vec{r}}$ and the electric gauge-field operators:
\begin{equation}
    \mathfrak{B}_{\text{gauge-inv}}
    =
    \left\langle
    Z_{\vec{r}} \mathcal{Z}^{(1)}_{\vec{r}+\frac{1}{2}(\hat{x}+\hat{y})}Z^\dagger_{\vec{r}+\hat{x}+\hat{y}},\,
    Z_{\vec{r}+\hat{x}}\mathcal{Z}^{(2)}_{\vec{r}+\frac{1}{2}(\hat{x}+\hat{y})}Z^\dagger_{\vec{r}+\hat{y}},\,
    X_{\vec{r}},\,
    \mathcal{X}^{(1)}_{\vec{r}+\frac{1}{2}(\hat{x}+\hat{y})},\,
    \mathcal{X}^{(2)}_{\vec{r}+\frac{1}{2}(\hat{x}+\hat{y})}
    \right\rangle.
    \label{bvbv}
\end{equation}

Within the physical Hilbert space, the Gauss-law constraint can equivalently be expressed as
\begin{equation}
X_{\vec{r}} =
\left(\mathcal{X}^{(1)}_{\vec{r}-\frac{1}{2}(\hat{x}+\hat{y})}\right)^\dagger\,
\mathcal{X}^{(1)}_{\vec{r}+\frac{1}{2}(\hat{x}+\hat{y})}
\left(\mathcal{X}^{(2)}_{\vec{r}-\frac{1}{2}(\hat{x}-\hat{y})}\right)^\dagger\,
\mathcal{X}^{(2)}_{\vec{r}+\frac{1}{2}(\hat{x}-\hat{y})}.
\label{Gaussss}
\end{equation}

\subsubsection{Plaquette Operators}

Let us construct the plaquette operators as the minimal Wilson loops $W^{(\bar\alpha)}(\gamma_{\Box})$. Taking
$\vec r$ as the base point, the corresponding sequence of bond labels is
\begin{equation}
\gamma_{\Box}
=
\{(\vec{r},a_1),(\vec{r}+\hat{x},a_2),
(\vec{r}+\hat{x}-\hat{y},a_3),(\vec{r}-\hat{y},a_4)\}.
\end{equation}
The flavor assignment that supports nontrivial solutions for the
coefficients $g$ is $a_1=a_3=1$ and $a_2=a_4=2$.
The empty-operator condition \eqref{empty_operator_condition} then reads
\begin{equation}
g_{\vec{r}}^{(1,\bar\alpha)} \Delta^{(1)}_{\vec{r},\vec{s}}+g_{\vec{r}+\hat{x}}^{(2,\bar\alpha)} \Delta^{(2)}_{\vec{r}+\hat{x},\vec{s}}+
g_{\vec{r}+\hat{x}-\hat{y}}^{(1,\bar\alpha)} \Delta^{(1)}_{\vec{r}+\hat{x}-\hat{y},\vec{s}}+
g_{\vec{r}-\hat{y}}^{(2,\bar\alpha)} \Delta^{(2)}_{\vec{r}-\hat{y},\vec{s}}=0, ~~\forall \vec{s}.
\end{equation}
Using Eq.~\eqref{zaq}, this becomes
\begin{eqnarray}
&&\delta_{\vec{r},\vec{s}}\left(g_{\vec{r}}^{(1,\bar\alpha)}+g_{\vec{r}-\hat{y}}^{(2,\bar\alpha)}\right)-
\delta_{\vec{r}+\hat{x}+\hat{y},\vec{s}}\left(g_{\vec{r}}^{(1,\bar\alpha)}-g_{\vec{r}+\hat{x}}^{(2,\bar\alpha)}\right)\nonumber\\
&-& \delta_{\vec{r}+2\hat{x},\vec{s}}\left(g_{\vec{r}+\hat{x}}^{(2,\bar\alpha)}+g_{\vec{r}+\hat{x}-\hat{y}}^{(1,\bar\alpha)}\right)+
\delta_{\vec{r}+\hat{x}-\hat{y},\vec{s}}\left(g_{\vec{r}+\hat{x}-\hat{y}}^{(1,\bar\alpha)}-g_{\vec{r}-\hat{y}}^{(2,\bar\alpha)}\right)=0.
\end{eqnarray}
Thus, the four coefficients obey three independent relations:
\begin{equation}
\begin{aligned}
g_{\vec{r}}^{(1,\bar\alpha)}
 +g_{\vec{r}-\hat{y}}^{(2,\bar\alpha)}&=0,\\
g_{\vec{r}}^{(1,\bar\alpha)}
 -g_{\vec{r}+\hat{x}}^{(2,\bar\alpha)}&=0,\\
g_{\vec{r}+\hat{x}}^{(2,\bar\alpha)}
 +g_{\vec{r}+\hat{x}-\hat{y}}^{(1,\bar\alpha)}&=0.
\end{aligned}
\label{setofconst}
\end{equation}
The fourth relation follows from these three.

For each $\bar\alpha$, Eq.~\eqref{setofconst} leaves one free
coefficient on the plaquette. Denoting it by
$c_{\vec r}^{(\bar\alpha)}
\equiv g_{\vec r}^{(1,\bar\alpha)}$, we obtain
\begin{equation}
g_{\vec r}^{(1,\bar\alpha)}
=g_{\vec r+\hat x}^{(2,\bar\alpha)}
=c_{\vec r}^{(\bar\alpha)},
\qquad
g_{\vec r-\hat y}^{(2,\bar\alpha)}
=g_{\vec r+\hat x-\hat y}^{(1,\bar\alpha)}
=-c_{\vec r}^{(\bar\alpha)}.
\label{local_g_solution}
\end{equation}
To determine how many independent assignments of these coefficients
exist across the lattice, we compare the solutions on translated
plaquettes.

First, consider a plaquette with base point $\vec r+\hat x$. Its
coefficients satisfy
\begin{equation}
\begin{aligned}
g_{\vec r+\hat x}^{(1,\bar\alpha)}
 +g_{\vec r+\hat x-\hat y}^{(2,\bar\alpha)}&=0,\\
g_{\vec r+\hat x}^{(1,\bar\alpha)}
 -g_{\vec r+2\hat x}^{(2,\bar\alpha)}&=0,\\
g_{\vec r+2\hat x}^{(2,\bar\alpha)}
 +g_{\vec r+2\hat x-\hat y}^{(1,\bar\alpha)}&=0.
\end{aligned}
\label{setofconst1}
\end{equation}
This plaquette shares no coefficient with the one based at $\vec r$.
Consequently, these two local equations impose no relation between
$c_{\vec r}^{(\bar\alpha)}$ and
$c_{\vec r+\hat x}^{(\bar\alpha)}$.

By contrast, a plaquette based at $\vec r+\hat x+\hat y$ satisfies
\begin{equation}
\begin{aligned}
g_{\vec r+\hat x+\hat y}^{(1,\bar\alpha)}
 +g_{\vec r+\hat x}^{(2,\bar\alpha)}&=0,\\
g_{\vec r+\hat x+\hat y}^{(1,\bar\alpha)}
 -g_{\vec r+2\hat x+\hat y}^{(2,\bar\alpha)}&=0,\\
g_{\vec r+2\hat x+\hat y}^{(2,\bar\alpha)}
 +g_{\vec r+2\hat x}^{(1,\bar\alpha)}&=0.
\end{aligned}
\label{setofconst2}
\end{equation}
It shares $g_{\vec r+\hat x}^{(2,\bar\alpha)}$ with the plaquette
based at $\vec r$. Equation~\eqref{local_g_solution} therefore fixes
its amplitude to be
$c_{\vec r+\hat x+\hat y}^{(\bar\alpha)}
=-c_{\vec r}^{(\bar\alpha)}$.
Applying the same argument to a plaquette based at
$\vec r+\hat x-\hat y$ gives
\begin{equation}
c_{\vec r+\hat x+\hat y}^{(\bar\alpha)}
=-c_{\vec r}^{(\bar\alpha)},
\qquad
c_{\vec r+\hat x-\hat y}^{(\bar\alpha)}
=-c_{\vec r}^{(\bar\alpha)}.
\label{plaquette_amplitude_recurrences}
\end{equation}


The shifts $\hat x+\hat y$ and $\hat x-\hat y$ connect all base
points of a given parity but do not connect the two parity
sublattices. Consequently, one initial amplitude determines all
amplitudes on each sublattice, while the two initial amplitudes
remain independent. The global coefficient assignments therefore
have two independent generators. We choose $g^{(a,1)}$ to be
nonzero on plaquettes with odd base points and to vanish on those
with even base points, and choose $g^{(a,2)}$ with the opposite
assignment. We normalize each generator by setting
$c_{\vec r}^{(\bar\alpha)}=1$ at a chosen base point of the
corresponding parity. The remaining amplitudes are fixed by
Eq.~\eqref{plaquette_amplitude_recurrences}. Up to inversion,
which leaves the Hermitian combination of plaquette terms unchanged, the
resulting operators take the form
\begin{equation}
W^{(\bar\alpha)}(\gamma_{\square}) =\mathcal{Z}^{(1)}_{\vec{r}+\frac{1}{2}(\hat{x}+\hat{y})}
\mathcal{Z}^{(2)}_{\vec{r}+\frac{1}{2}(3\hat{x}+\hat{y})}
\left(\mathcal{Z}^{(1)}_{\vec{r}+\frac{1}{2}(3\hat{x}-\hat{y})}\right)^{\dagger}
\left(\mathcal{Z}^{(2)}_{\vec{r}+\frac{1}{2}(\hat{x}-\hat{y})}\right)^{\dagger},
\label{plaquettess}
\end{equation}
where $\vec r$ is odd for $\bar\alpha=1$ and even for
$\bar\alpha=2$.


\subsubsection{Extended Operators}

We now discuss extended Wilson operators constructed from the minimal
plaquette operators $W^{(\bar\alpha)}(\gamma_{\Box})$. Consider an
oriented closed sequence of original-lattice vertices,
\begin{equation}
\gamma = \{(\vec{r}_1,a_1),\ldots,(\vec{r}_n,a_n)\},
\qquad
\vec r_{n+1}\equiv\vec r_1,
\end{equation}
where all $\vec r_i$ have the same parity and consecutive vertices
are joined by diagonal segments. For each segment, we assign
$a_i=1$ if it lies along an ascending diagonal and $a_i=2$ if it
lies along a descending diagonal. The corresponding gauge operator
is located at the midpoint
\begin{equation}
\vec r_{a_i}
=
\frac{1}{2}(\vec r_i+\vec r_{i+1}).
\label{path009}
\end{equation}
Thus the vertices $\vec r_i$ specify the path, while the pair
consisting of a segment midpoint and its flavor specifies the
gauge operator appearing in the Wilson loop.

The exponent of each gauge operator is fixed by the direction in
which the sequence traverses its segment. We orient both types of
diagonal segment toward increasing $x$ and define
\begin{equation}
\sigma_i
=
\operatorname{sgn}\!\left[
(\vec r_{i+1}-\vec r_i)\cdot\hat x
\right].
\label{wilson_sign_rule}
\end{equation}
A segment traversed toward increasing $x$ therefore has exponent
$+1$, while one traversed toward decreasing $x$ has exponent $-1$.
For the elementary loop, the four segments are traversed twice to
the right and then twice to the left, reproducing the coefficient
pattern $(+1,+1,-1,-1)$ obtained from
Eq.~\eqref{setofconst}.

\begin{figure}
    \centering

    \begin{subfigure}{0.50\textwidth}
        \centering
        \includegraphics[width=\linewidth]{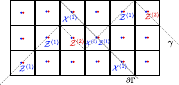}
        \caption{}
        \label{fig15}
    \end{subfigure}
    \hfill
    \begin{subfigure}{0.48\textwidth}
        \centering
        \includegraphics[width=\linewidth]{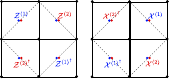}
        \caption{}
        \label{fig10}
    \end{subfigure}
    \caption{(a) 1-form symmetry operator $W^{(\alpha)}(\gamma)$ composed of $\mathcal{Z}^{(i)}$ operators, and $U^{(\alpha)}_{\partial}(\partial\Gamma)$ composed of $\mathcal{X}^{(i)}$ operators. The paths $\gamma$ and $\partial\Gamma$ are illustrated by the dotted lines along the diagonals and the operators act on each site of the dual lattices that the paths intersect. These operators intersect at a point on the lattice where they do not commute. (b) Bond operators in Eq. (\ref{bond615}). These local operators commute with the operators in (a).}
    \label{fig:1}
\end{figure}

With these assignments, the Wilson operators in Eq.~\eqref{Wilson}
take the form
\begin{equation}
\begin{aligned}
W^{(1)}(\gamma_{\mathrm{odd}})
&=
\prod_{i=1}^{n}
\left(
\mathcal Z^{(a_i)}_{\frac12(\vec r_i+\vec r_{i+1})}
\right)^{\sigma_i},\\
W^{(2)}(\gamma_{\mathrm{even}})
&=
\prod_{i=1}^{n}
\left(
\mathcal Z^{(a_i)}_{\frac12(\vec r_i+\vec r_{i+1})}
\right)^{\sigma_i}.
\end{aligned}
\label{rty1}
\end{equation}
Here $\gamma_{\mathrm{odd}}$ and $\gamma_{\mathrm{even}}$ denote
sequences whose vertices $\vec r_i$ belong to the odd and even
sublattices, respectively. For a contractible sequence, the
resulting Wilson loop can be obtained by assembling the minimal
plaquette operators enclosed by $\gamma$; contributions on
interior segments cancel, leaving precisely the factors along
the sequence. An example is shown in Fig.~\ref{fig15}.

We next construct the boundary operators defined in Eq.~\eqref{eqU}.
In $d=2$, the boundary of a two-dimensional region $\Gamma$ is a
closed loop, which we denote by $\partial\Gamma\equiv\tilde\gamma$.
We traverse the outer boundary $\tilde\gamma$ clockwise, so that
the interior of $\Gamma$ lies to the right of each oriented segment.
As for the Wilson loops, we write $\tilde\gamma_{\mathrm{odd}}$
or $\tilde\gamma_{\mathrm{even}}$ according to the parity of its
original-lattice vertices $\vec r_i$.

The flavor assignment is opposite to that used for the Wilson
loops: $b_i=1$ when the segment from $\vec r_i$ to
$\vec r_{i+1}$ lies along a descending diagonal, and $b_i=2$
when it lies along an ascending diagonal. The associated gauge
operator is located at
$\frac12(\vec r_i+\vec r_{i+1})$. Its exponent follows directly
from the coefficients $\Delta^{(b_i)}$ in Eq.~\eqref{zaq}.
With the orientation specified above, it is
\begin{equation}
\tilde\sigma_i
=
\begin{cases}
\phantom{-}\operatorname{sgn}\!\left[
(\vec r_{i+1}-\vec r_i)\cdot\hat x
\right], & \text{if}~~b_i=1,\\[2pt]
-\operatorname{sgn}\!\left[
(\vec r_{i+1}-\vec r_i)\cdot\hat x
\right], & \text{if}~~b_i=2.
\end{cases}
\label{boundary_sign_rule}
\end{equation}
The boundary operators therefore take the form
\begin{equation}
\begin{aligned}
U^{(1)}_{\partial}(\tilde\gamma_{\mathrm{even}})
&=
\prod_{i=1}^{n}
\left(
\mathcal X^{(b_i)}_{\frac12(\vec r_i+\vec r_{i+1})}
\right)^{\tilde\sigma_i},\\
U^{(2)}_{\partial}(\tilde\gamma_{\mathrm{odd}})
&=
\prod_{i=1}^{n}
\left(
\mathcal X^{(b_i)}_{\frac12(\vec r_i+\vec r_{i+1})}
\right)^{\tilde\sigma_i}.
\end{aligned}
\label{rty2}
\end{equation}




Consider two noncontractible loops $\gamma$ and $\tilde{\gamma}$ that have a single intersection. The corresponding line operators obey the nontrivial commutation relations
\begin{eqnarray}
 W^{(1)}(\gamma_{\text{odd}}) U^{(1)}_{\partial}(\tilde{\gamma}_{\text{even}})&=&\omega^{\ast} U^{(1)}_{\partial}(\tilde{\gamma}_{\text{even}})W^{(1)}(\gamma_{\text{odd}});\nonumber\\
 W^{(2)}(\gamma_{\text{even}}) U^{(2)}_{\partial}(\tilde{\gamma}_{\text{odd}})&=&\omega^{\ast} U^{(2)}_{\partial}(\tilde{\gamma}_{\text{odd}}) W^{(2)}(\gamma_{\text{even}}).
  \label{dfs}
\end{eqnarray}
An explicit configuration of line operators realizing these commutation relations is illustrated in Fig.~\ref{fig15}. 

We now consider the subalgebra of \eqref{bvbv} for which the extended
operators \eqref{rty1} and \eqref{rty2} are symmetries. Following
Eq.~\eqref{nnmm}, the resulting pure-gauge bond algebra is generated
by the gauge-field parts of the Gauss-law operators
\eqref{Gaussss} and the plaquette operators \eqref{plaquettess}:
\begin{eqnarray}
\mathfrak{B}_{\text{pure-gauge}}
&=&
\Bigl\langle
\left(\mathcal{Z}^{(2)}_{\vec{r}-\frac{1}{2}(\hat{x}+\hat{y})}\right)^{\dagger}
\mathcal{Z}^{(2)}_{\vec{r}+\frac{1}{2}(\hat{x}+\hat{y})}
\mathcal{Z}^{(1)}_{\vec{r}-\frac{1}{2}(\hat{x}-\hat{y})}
\left(\mathcal{Z}^{(1)}_{\vec{r}+\frac{1}{2}(\hat{x}-\hat{y})}\right)^{\dagger},
\nonumber\\
&&
\left(\mathcal{X}^{(1)}_{\vec{r}-\frac{1}{2}(\hat{x}+\hat{y})}\right)^{\dagger}
\mathcal{X}^{(1)}_{\vec{r}+\frac{1}{2}(\hat{x}+\hat{y})}
\left(\mathcal{X}^{(2)}_{\vec{r}-\frac{1}{2}(\hat{x}-\hat{y})}\right)^{\dagger}
\mathcal{X}^{(2)}_{\vec{r}+\frac{1}{2}(\hat{x}-\hat{y})}
\Bigr\rangle.
\label{bond615}
\end{eqnarray}
Here the plaquette Wilson operators have been written with a shifted
base point relative to Eq.~\eqref{plaquettess}. The generators of
this bond algebra are illustrated in Fig.~\ref{fig10}.


A representative Hamiltonian constructed from the pure-gauge bond algebra \eqref{bond615} is
\begin{eqnarray}
H_{\text{pure-gauge}}
&=&
-J_{\mathrm{m}}\sum_{\vec{r}}
\left(\mathcal{Z}^{(2)}_{\vec{r}-\frac{1}{2}(\hat{x}+\hat{y})}\right)^{\dagger}
\mathcal{Z}^{(2)}_{\vec{r}+\frac{1}{2}(\hat{x}+\hat{y})}
\mathcal{Z}^{(1)}_{\vec{r}-\frac{1}{2}(\hat{x}-\hat{y})}
\left(\mathcal{Z}^{(1)}_{\vec{r}+\frac{1}{2}(\hat{x}-\hat{y})}\right)^{\dagger}
\nonumber\\
&&
-J_{\mathrm{e}}\sum_{\vec{r}}
\left(\mathcal{X}^{(1)}_{\vec{r}-\frac{1}{2}(\hat{x}+\hat{y})}\right)^{\dagger}
\mathcal{X}^{(1)}_{\vec{r}+\frac{1}{2}(\hat{x}+\hat{y})}
\left(\mathcal{X}^{(2)}_{\vec{r}-\frac{1}{2}(\hat{x}-\hat{y})}\right)^{\dagger}
\mathcal{X}^{(2)}_{\vec{r}+\frac{1}{2}(\hat{x}-\hat{y})}
+\mathrm{H.c.},
\label{eq:pure_gauge_Hamiltonian}
\end{eqnarray}
where $J_{\mathrm{m}}$ and $J_{\mathrm{e}}$ control the magnetic and electric terms, respectively. The ground-state manifold of this Hamiltonian realizes the topologically ordered phase associated with $\mathcal{D}(\mathbb{Z}_N\times\mathbb{Z}_N)$. Its generalized symmetry structure is
\begin{eqnarray}
G^{\vee}
=
\left(
(\mathbb{Z}_N\times\mathbb{Z}_N)^{(1)}
\rtimes
(\mathbb{Z}_N\times\mathbb{Z}_N)^{(1)}
\right)
\rtimes_{\varphi^\vee}
G_{\mathrm{space}},
\end{eqnarray}
with the corresponding 't Hooft anomalies given in \eqref{dfs}. The nontrivial action $\varphi^\vee$ of the space group on the one-form symmetries therefore endows this topologically ordered phase with the structure of an isotropic SET.


\subsection{Symmetry-Enriched Topological Order}

The elementary anyonic excitations are created at the endpoints of the open-string versions of the operators in Eqs.~(\ref{rty1}) and (\ref{rty2}). In particular, an open Wilson operator $W^{(\bar\alpha)}$ creates a pair of electric anyons $e_{\bar\alpha}$, whereas an open operator $U_{\partial}^{(\alpha)}$ creates a pair of magnetic anyons $m_{\alpha}$. Since $\alpha, \bar\alpha=1,2$, the theory contains four elementary anyonic generators, $e_1$, $e_2$, $m_1$, and $m_2$, whose fusion products generate the full set of anyon types of $\mathcal{D}(\mathbb{Z}_N\times\mathbb{Z}_N)$.

To determine the action of the space group on the anyons in the present example, we first recall the general transformation rules derived in Eqs.~\eqref{eq: flux anyons transformation} and \eqref{eq: charge anyons transformation}:
\begin{equation}
g: m_{\alpha}
\rightarrow
\prod_{\beta}
\left(m_\beta\right)^{\left(A^{(g)}\right)^{-T}_{\alpha,\beta}},
\qquad
g: e_{\bar\alpha}
\rightarrow
\prod_{\bar\beta}
\left(e_{\bar\beta}\right)^{A^{(g)}_{\bar\alpha,\bar\beta}}.
\end{equation}
For the primitive translations $g=T_x$ and $T_y$, the representation matrices satisfy $A^{(T_x)}=A^{(T_y)}$, as in \eqref{bnbn}, and this common matrix is symmetric and obeys $A^{-1}=A$. It follows that $\left(A^{(g)}\right)^{-T}=A^{(g)}$, so the magnetic representation and the contragredient electric representation coincide. Substituting the explicit translation matrices into the general transformation rules gives
\begin{equation}
g:
\begin{cases}
~e_1 \rightarrow e_2, \quad ~~~e_2 \rightarrow e_1,\\
m_1 \rightarrow m_2, \quad m_2 \rightarrow m_1.
\end{cases}
\qquad g=T_x,T_y.
\end{equation}
Thus, either primitive translation exchanges the labels $1$ and $2$ in both the electric and magnetic sectors. This permutation reflects the association of the anyons labeled by $\alpha, \bar\alpha=1$ and $\alpha,\bar\alpha=2$ with the odd and even sublattices, respectively, which are exchanged by a single lattice translation. By contrast, since $A^{(R)}=1$, the general transformation rules imply that the rotation $R$ leaves each elementary anyon invariant.



\section{Conclusions and Outlook}

In this work, we developed a constructive framework for discrete Abelian modulated symmetries and their gauging in $d$-dimensional lattice systems. The modulation profiles were organized as a finite-dimensional representation of the space group acting on the internal symmetry generators. For a finite internal symmetry group, this action necessarily factors through a finite quotient of the space group, where the nontrivial spatial dependence of the symmetry is therefore controlled by finite-order transformations. This allows for an extensive classification scheme of such modulating symmetries once $G_{\mathrm{int}}$ is fixed.

A central element of our construction is the systematic determination of the local symmetric bond algebra. For a chosen finite size stencil, the symmetry constraints are encoded in a matrix built directly from the modulation functions, while the allowed local interactions are obtained from its kernel. The same construction also naturally gives rise to an algebra of open operators, where products of local bonds that are empty along closed sequences acquire nontrivial boundary contributions when the sequences are opened, and these boundary operators have a nontrivial algebra with truncated symmetry operators.

Gauging the internal symmetry reorganizes this operator algebra. The original global symmetry operators become trivial on the physical Hilbert space, whereas the empty operators of the ungauged theory are promoted to nontrivial Wilson operators generating emergent modulated higher-form quantum symmetries. Conversely, truncated versions of the original symmetry operators become nontrivial boundary operators.

In $(2+1)$ dimensions, this construction provides a direct microscopic route from modulated symmetries to Abelian symmetry-enriched topological phases. For the orientation-preserving spatial symmetries considered here,  the modulation profiles completely determine the corresponding permutation of the anyons.
The electric and magnetic sectors transform in mutually contragradient representations that preserve  their canonical braiding pairing. This identifies the modulation data of the ungauged theory with the anyon-permutation data of the gauged SET phase.

Although the approach taken in this work is somewhat general, several directions to make it even more comprehensive remain open. To start, our construction assigns only one degree of freedom per site, while allowing multiple on-site species could lead to potentially richer modulation patterns.  Similarly, after gauging, allowing for more general transformations among gauge-field flavors than just monomial transformations could potentially enlarge the class of spatial actions accessible within this framework.

The construction could also be enriched by stacking the system with
$G_{\mathrm{space}}$ SPT phases, providing a route to a broader class
of SETs. Further extensions include microscopic realizations in
which spatial symmetries exchange electric and magnetic anyons, as
well as those involving orientation-reversing crystalline symmetries.
Examples of both appear in the literature, but lie beyond the present
framework. Finally, the modulated mixed anomaly identified here may
constrain transitions between deconfined, confined, and
symmetry-breaking phases, which still remain largely unexplored.  Following up on these directions could clarify even further
how spatially modulated symmetries can be used to organize phases of matter and transitions between them.
\begin{acknowledgments}
We thank Salvatore Pace for insightful discussions during the early stages of this work.	The authors acknowledge the use of ChatGPT (OpenAI) for helpful discussions and assistance in improving the clarity and presentation of the manuscript. We acknowledge the financial support from the Brazilian funding agencies Fundação Araucária, CAPES, and CNPq.  {The work of G.D. is supported by the U.S. Department of Energy, Office of Basic Energy Sciences, under Award DE-SC0026189.} 
	
\end{acknowledgments}


\bibliographystyle{unsrt}
\bibliography{ref}

\end{document}